\documentclass[journal]{IEEEtran}
\usepackage{xcolor} % Required for colors

\newcommand{\rev}[1]{\textcolor{black}{#1}}

\usepackage{cite}
\usepackage{subcaption}
\usepackage{float}
\usepackage{hyperref}
\usepackage{graphicx}
\usepackage[dvipsnames, x11names,table]{xcolor}
\newcommand{\defcommenter}[2]{%
  \expandafter\newcommand\csname #1\endcsname[1]{%
  {\color{#2}[#1: ##1]}%
  }%
}
\defcommenter{TODO}{blue}
\usepackage{amsmath,amssymb,amsfonts}
\usepackage{array}
\newcolumntype{L}[1]{>{\raggedright\let\newline\\\arraybackslash\hspace{0pt}}m{#1}}
\newcolumntype{C}[1]{>{\centering\let\newline\\\arraybackslash\hspace{0pt}}m{#1}}
\newcolumntype{R}[1]{>{\raggedleft\let\newline\\\arraybackslash\hspace{0pt}}m{#1}}
\usepackage{breqn} %  Break lines in equations

\usepackage[font=small]{caption}

\usepackage{amsmath}

\usepackage{algorithmic}
\usepackage{array}

\usepackage{url}

\usepackage{etoolbox}

\usepackage{eso-pic}

\newcommand{\FirstPageHeaderNotice}{%
  \AddToShipoutPictureFG*{%
    \AtPageUpperLeft{%
      \raisebox{-12mm}{% <-- vertical offset from top edge (tune)
        \hspace*{18mm}% <-- left offset from left edge (tune)
        \parbox{\dimexpr\paperwidth-36mm\relax}{\scriptsize
          \copyright~2026 IEEE. This manuscript has been accepted to IEEE Transactions on Circuits and Systems I:Regular Papers. Personal use of this material is permitted.  Permission from IEEE must be obtained for all other uses, in any current or future media, including reprinting/republishing this material for advertising or promotional purposes, creating new collective works, for resale or redistribution to servers or lists, or reuse of any copyrighted component of this work in other works.
        }%
      }%
    }%
  }%
}

\begin{document}
\FirstPageHeaderNotice

\title{A Reconfigurable Time-Domain In-Memory Computing Macro using FeFET-Based CAM with Multilevel Delay Calibration in 28 nm CMOS}

\author{Jeries~Mattar,~\IEEEmembership{Student Member,~IEEE,}
        Mor M.~Dahan,~\IEEEmembership{Student Member,~IEEE,}
        Stefan~D\"unkel,
        Halid~Mulaosmanovic,~\IEEEmembership{Senior Member,~IEEE,}
        Gunda Beernink,
        Sven~Beyer,         Eilam~Yalon,~\IEEEmembership{Member,~IEEE},~and~Nicolás~Wainstein,~\IEEEmembership{Member,~IEEE}% <-this % stops a space
\thanks{J. Mattar, M. M. Dahan, E. Yalon, and N. Wainstein are with the Andrew and Erna Viterbi Faculty of Electrical and Computer Engineering, Technion - Israel Institute of Technology, Haifa, Israel. Corresponding author: e-mail: (nicolasw@technion.ac.il).}% <-this % stops a space
\thanks{S. D\"unkel, H. Mulaosmanovic, G. Beernink, and S. Beyer are with GlobalFoundries Fab1 LLC and Co. KG, Dresden, Germany.}
\thanks{This work is funded by the European Union within "NextGeneration EU", by the Federal Ministry for Economic Affairs and Climate Action (BMWK) on the basis of a decision by the German Bundestag, by the State of Saxony with tax revenues based on the budget approved by the members of the Saxon State Parliament in the framework of “Important Project of Common European Interest - Microelectronics and Communication Technologies", under the project name “EUROFOUNDRY”, and by the Uzia Galil Memorial Fund.
 }
\thanks{Manuscript received XX, 2025; revised XX, 2025.}}

\maketitle

\begin{abstract}
Time-domain nonvolatile in-memory computing (TD-nvIMC) offers a promising pathway to reduce data movement and improve energy efficiency by encoding computation in delay rather than voltage or current. This work presents a fully integrated and reconfigurable TD-nvIMC macro, fabricated in 28~nm CMOS, that combines a ferroelectric FET (FeFET)-based content-addressable memory array, a cascaded delay element chain, and a time-to-digital converter. The architecture supports binary multiply-and-accumulate (MAC) operations using XOR- and AND-based matching, as well as in-memory Boolean logic and arithmetic functions. Sub-nanosecond MAC resolution is achieved through experimentally demonstrated 550~ps delay steps, representing a 2000$\times$ improvement over prior FeFET TD-nvIMC work, enabled by multilevel-state calibration with $\leq$~100~ps resolution. Write-disturb resilience is ensured via isolated triple-well bulks. The proposed macro achieves a measured throughput of 222.2~MOPS/cell and energy efficiency of 1887 TOPS/W at 0.85~V, establishing a viable path toward scalable, energy-efficient TD-nvIMC accelerators.

\end{abstract}


% \fi
\IEEEpeerreviewmaketitle

\section{Introduction}
\IEEEPARstart{A}{rtificial} neural networks (ANN) have demonstrated exceptional performance in various machine learning tasks, such as natural language processing, speech recognition, and image classification~\cite{Murmann2021,Sze2017}. ANN hardware accelerators rely on highly parallel multiply-and-accumulate (MAC) operations, generating significant intermediate data that must be frequently transferred between the processing element (PE) and memory in conventional von Neumann architectures. This results in limited energy efficiency and significant latency, especially for complex models with high bit precision, limiting their use on the edge~\cite{ShimengAug2021,Sebastian2020nat,ShimengApr2021,Jiang2020,Chuan2021}.

\begin{figure*}[!t]
    \centering
    \includegraphics[width=0.9\textwidth,trim={0.6cm 0.6cm 0.6cm 0.6cm},clip]{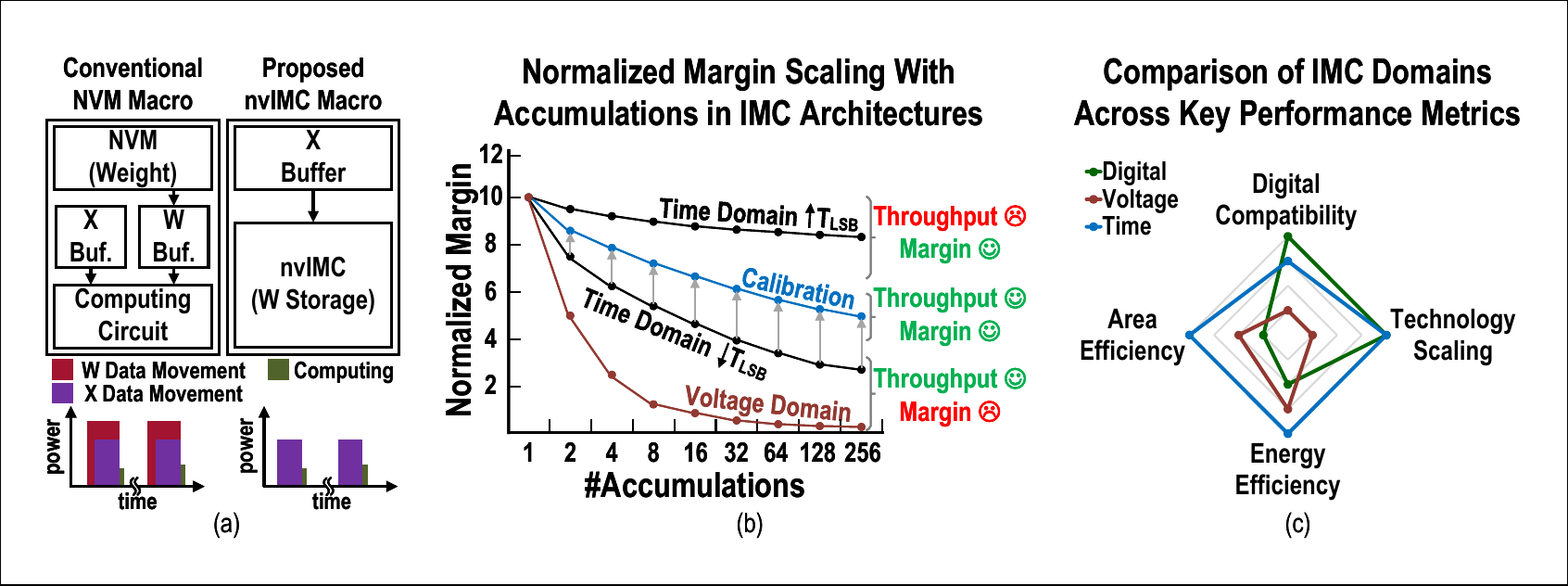} 
    \caption{ Challenges in nvIMC. (a) Comparison between conventional NVM and proposed nvIMC macro showing reduced data movement and power. (b) Normalized margin comparison showing improved throughput and enhanced noise margins in time-domain via calibration. (c) Radar chart benchmarking digital, voltage-domain, and time-domain IMC approaches across key performance metrics.}
    \label{fig:intro}
\end{figure*}

Nonvolatile in-memory-computing (nvIMC) architectures, leveraging nonvolatile memories (NVMs) can integrate computation and memory, minimizing data movement and improving energy efficiency [see Fig.~\ref{fig:intro}(a)]. Analog-based nvIMCs~\cite{Ielmini2018,LeGallo2023} enable in-memory MAC operations but face challenges such as noise susceptibility, device-to-device (D2D) variations, and reliance on power- and area-inefficient data converters. To address these challenges, time-domain nonvolatile IMC (TD-nvIMC) performs MAC operations by accumulating time delays through a chain of delay elements (DEs), where each delay is modulated by input activations and stored weights. The resulting phase difference is then digitized using a compact, standard cell-based time-to-digital converter (TDC), enabling improved energy efficiency and scalability.

% To address these challenges, time-domain nvIMC (TD-nvIMC) uses cumulative delay of cascaded delay elements (DEs) modulated by input activations and weights for MAC operations. The phase difference is then quantized by a compact standard cell-based time-to-digital converter (TDC), improving energy efficiency and scalability. 

TD-nvIMC offers improved signal margin over voltage-domain approaches, particularly when performing a large number of accumulations, as illustrated in Fig.~\ref{fig:intro}(b). In TD-nvIMC, decreasing the time resolution or least significant bit duration ($T_{LSB}$) improves both latency and throughput. However, this reduction also narrows the timing margin, making the system more susceptible to noise and variations\rev{, approaching the reduced signal margins typical of voltage-domain implementations}. To sustain high performance while preserving signal integrity, calibration techniques are essential to compensate for device mismatch and ensure reliable operation at fine time resolutions [see Fig.~\ref{fig:intro}(b)]. As qualitatively illustrated in Fig.~\ref{fig:intro}(c), TD-nvIMC achieves a favorable trade-off across key performance metrics, such as area efficiency, scalability, digital compatibility and energy efficiency, compared to digital and voltage-domain IMC architectures.

Ferroelectric FET (FeFET) is a promising NVM for TD-nvIMC, due to its large memory window, fast and low write power, and high endurance~\cite{Mikolajick2020}. Prior works on FeFET-based TD-nvIMC demonstrated early integration of FeFETs for delay modulation and weight storage~\cite{Luo2021,Yin2024,Rafiq2023}. A DE based on a current-starved inverter (CSI) integrating a Fe-FinFET~\cite{Luo2021} experimentally demonstrated delay steps and total accumulated delay $(T_D)>10~\mu s$. A FeFET array featuring two devices per cell is used to switch a load capacitor in an inverter chain~\cite{Yin2024}, demonstrating Boolean XOR and AND logic operations. A $T_D>1~\mu s$ is experimentally demonstrated using discrete devices on a breadboard, with time steps $\sim 1~\mu s$. 
Although these works pioneered the use of FeFET for TD-nvIMC, they primarily focused on individual device demonstrations, lacked a fully integrated macro and exhibited large delays and latency~($>1~\mu s$). Furthermore, prior implementations do not address key challenges such as D2D and timing mismatches, nor do they incorporate critical components required for a functional TD-nvIMC macro, such as a fully integrated DE and TDC. 

In this work, we present a reconfigurable FeFET-based TD-nvIMC macro with integrated content-addressable memory (CAM), DE chain, and time-to-digital converter (TDC). Fabricated in a 28 nm CMOS process, our reconfigurable macro supports binary XOR- and AND-based MAC operations, as well as logic (AND, OR) and full adder. A key novelty is a delay mismatch calibration scheme via multilevel state (MLS) with $\leq$100~ps temporal resolution ($\Delta t$). We present for the first time write-disturb prevention and MLS programming at array level using the bulk in a triple-well process. Furthermore, we experimentally demonstrate MAC operations with 550~ps delay steps ($\Delta s$), a 2000$\times$ improvement over~\cite{Luo2021,Yin2024}, and energy efficiency of 1887~TOPS/W.

The remainder of this article is organized as follows. Section~\ref{sec:background} introduces FeFET technology and its use in IMC, as well as time-domain IMC (TD-IMC) using both SRAM and NVMs. The memory array design, write-disturb prevention scheme and MLS using bulk are introduced in Section~\ref{sec:memory}. The proposed TD-nvIMC macro is introduced in Section~\ref{sec:arch}, while the reconfigurable IMC logic topologies are described in Section~\ref{sec:imc}. Then, the calibration scheme is presented in Section~\ref{sec:calibration} and the experimental results are presented in Section~\ref{sec:results}. A comparison with recent works is discussed in Section~\ref{sec:comparison}. Finally, the conclusions are presented in~\ref{sec:conclusion}.

% \begin{figure}
%     \centering
%     \includegraphics[width=1\columnwidth]{intro2.png}
%     \caption{Challenges in IMC.}
%     \label{fig:intro}
% \end{figure}

% \begin{figure}[!t]
%     \centering
%     \includegraphics[width=0.8\columnwidth,trim={0.6cm 0.6cm 0.6cm 0.6cm},clip]{figures/intro_table.pdf}
%     \caption{Comparison between IMC domains showing the advantages of TD-nvIMC. }
%     \label{fig:intro}
% \end{figure}

\section{Fundamentals of Ferroelectric FETs and Time-Domain In-Memory Computing}
\label{sec:background}
\subsection{Ferroelectric FET}
FeFET stands as a promising NVM for TD-IMC thanks to its high memory window, fast and low write power, and high endurance. In FeFETs, a ferroelectric layer is integrated into the gate stack of a field effect transistor. This ferroelectric layer presents two stable polarization ($P$) states at zero electrical field ($E$) that can be switched from one value to the other by applying an $E>E_C$, where $E_C$ is the coercive field ($E$ at which the effective $P$ is zero)~\cite{Mikolajick2020}. A ferroelectric material can be characterized by the hysteresis curve of $P$ or the displacement field $D$ as a function of $E$. 

A change in $P$ effectively results in a change in the threshold voltage ($V_{T}$) of the transistor, which varies between low to high threshold voltage (LVT and HVT, respectively). Since the demonstration of ferroelectric HfO\textsubscript{2}~\cite{Boscke2011}, the FeFET has been successfully integrated in a foundry CMOS process like 28~nm~\cite{Trentzsch2016}.
Unlike charge-based NVMs, FeFETs operate through E-induced polarization switching, which allows for low-energy write operations without high current flow. The read operation is performed at sub-coercive voltages, enabling non-destructive access and fast sensing of the stored $V_{T}$.

% The gate stack typically includes a thin interfacial SiO\textsubscript{2} and a ferroelectric doped HfO\textsubscript{2} layer to stabilize orthorhombic ferroelectric phases compatible with CMOS processes. These structural and material advantages have enabled FeFETs to demonstrate sub-nanosecond switching, high endurance, and long retention under aggressive cycling, with integration already shown at 28~nm nodes.

Beyond binary storage, FeFETs can support intermediate polarization states through partial switching of the ferroelectric domains. This property enables the realization of multi-level states (MLS)\cite{Mulaosmanovic2015}, where $V_{T}$ can be tuned continuously within the available memory window. Such MLS capabilities are highly valuable for applications requiring fine control over device behavior, including analog computing and calibration techniques. 
In addition to storage, this analog tunability positions FeFETs as a suitable candidate for time-domain computing and neuromorphic systems, where graded $V_{T}$ values can encode weights or control timing paths.

% Despite their advantages, FeFETs face known challenges such as polarization fatigue, charge trapping, and variability in $V_{T}$ due to defects or interface traps~\TODO{cite}. However, ongoing device engineering continues to improve reliability and scalability, bringing FeFETs closer to wide adoption in both memory and compute-in-memory architectures.

To simulate FeFET behavior, we adopt a hybrid modeling approach: (a) a dynamic model based on the Preisach framework~\cite{Dragosits2000} to capture program and erase behavior, and (b) a static model for steady-state analysis. In the Preisach-based model, ferroelectric polarization is represented as the collective response of a superposition of independent, non-interacting dipole domains within the ferroelectric layer. This polarization model is integrated with the NMOS model from the process design kit, enabling accurate reproduction of FeFET switching characteristics.
However, due to convergence issues and the inability to simulate initial conditions, the Preisach model is unsuitable for steady-state simulations. To address this, a simplified static model is employed, consisting of an ideal voltage source and a capacitor connected to the gate of an NMOS device, which emulates the FeFET behavior under high- and low-threshold voltage (HVT/LVT) conditions. This approach captures the I–V characteristics relevant to in-memory operations. All model parameters are calibrated against experimental data to ensure an accurate representation of FeFET behavior.

% To simulate the FeFET behavior, we use two approaches, a) the Preisach FeFET model~\TODO{cite} for dynamic behavior (\textit{i.e.}, program and erase), and b) a static model. In the Preisach model, the ferroelectric polarization is modeled as the collective response of a superposition of independent, non-interacting dipole domains within the ferroelectric layer. This ferroelectric model is integrated with the process design kit (PDK) NMOS model to accurately reproduce FeFET switching characteristics. Because of the convergence issues of the Preisach model and the inability to simulate the model with an initial condition, we use a static model to reproduce the static operation using an ideal voltage source combined with an NMOS. This provides the I-V characteristic of the FeFET in HVT and LVT. 
% Model parameters are calibrated against experimental measurements to ensure fidelity with observed device behavior. 

\subsection{Time-Domain In-Memory Computing}

\begin{figure}[!t]
    \centering
    \includegraphics[width=\columnwidth,trim={0.6cm 0.6cm 0.6cm 0.6cm},clip]{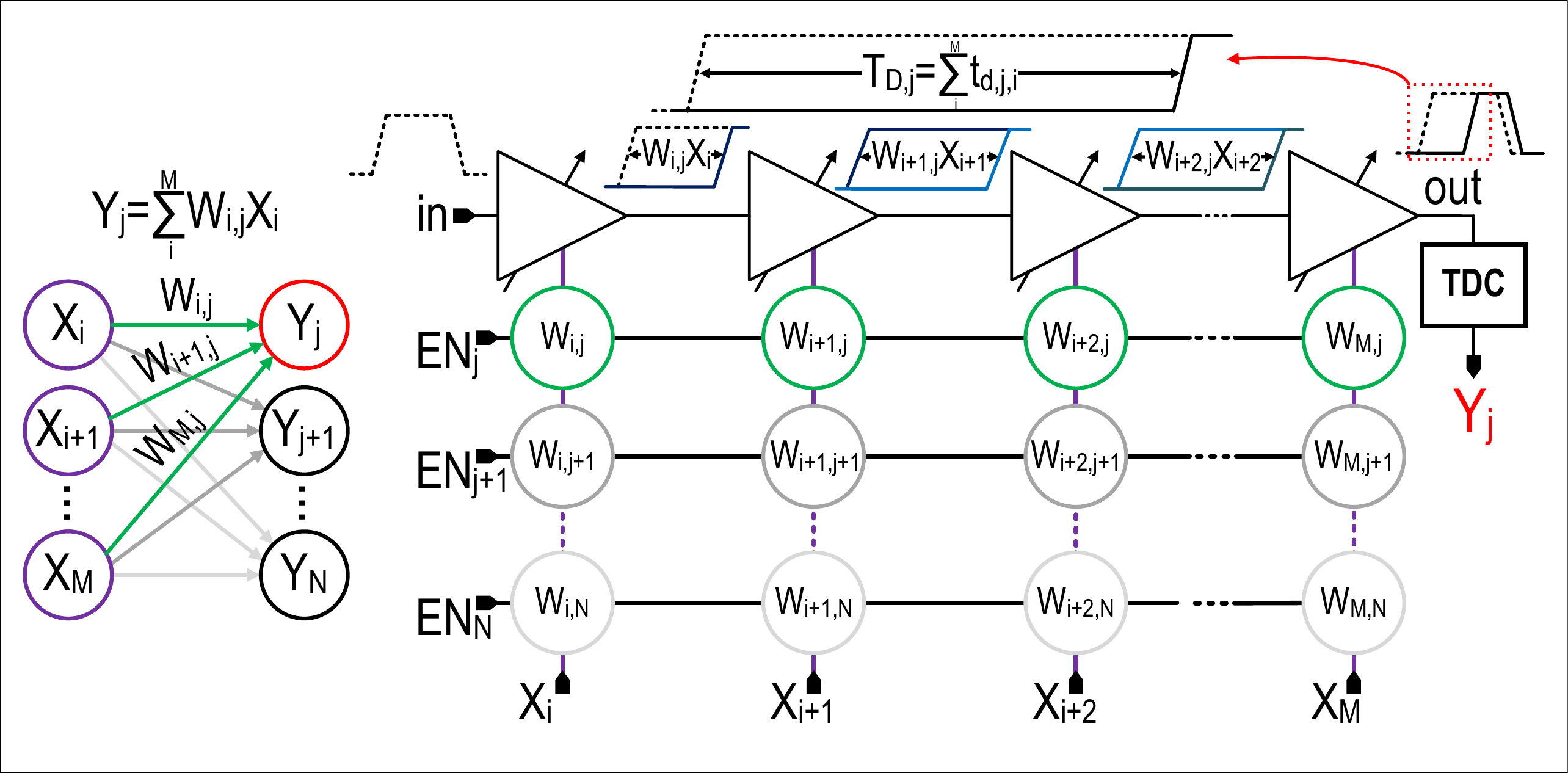}
    \caption{Schematic diagram of TD-nvIMC. The accumulated delay at the output is the sum of the stage delay, whose delay is proportional to $\sum_i W_{i,j}X_i$ for the enabled (EN) row.}
    \label{fig:tdimc_intro}
\end{figure}

Time-domain IMC utilizes DEs, whose propagation delay ($t_d$) is modulated proportionally to input activations ($X_i$) and stored weights ($W_{i,j}$), providing the multiply operation. As shown in Fig.~\ref{fig:tdimc_intro}, stage delays ($t_{d,i}$) gradually accumulate in a DE chain, providing the accumulation function. Previous works on SRAM-based TD-IMC architectures~\cite{Song2021,Yang2021} have demonstrated notable gains in energy efficiency and parallelism compared to traditional digital designs. In these architectures, each DE typically integrates a CSI controlled by binary weights stored in six-transistors (6T) SRAM cells. However, this approach incurs substantial area overhead, as each weight bit requires a full 6T cell along with associated analog control circuitry, limiting scalability and area density.

To improve density, reduce data movement, and minimize memory access overhead, recent research has investigated TD-IMC architectures leveraging NVMs~\cite{Luo2021,Yin2024,Seungchul2022,Zhang2021,Bavandpour2019,Sahay2020,Hung2023}. However, most TD-IMC implementations based on resistive memories, such as resistive RAM (ReRAM), phase-change memory (PCM), and spin-transfer torque MRAM (STT-MRAM) do not support direct delay modulation due to their inherently nonlinear and high ON resistance characteristics. For example,~\cite{Seungchul2022,Zhang2021} extracts timing information from the RC charging or discharging behavior of NVM cells, where the time required to reach a reference voltage depends on the programmed resistance and on the activation signal, which selects whether the cell participates in the operation. This approach avoids additional conversion stages but still suffers from poor linearity and limited timing resolution due to the exponential nature of the RC response and small resistance contrast. 

Similar charge/discharge-based concepts are employed in\cite{Bavandpour2019,Sahay2020}, using NOR Flash and ReRAM, respectively. These works attempt to improve timing linearity but still face scalability challenges. In\cite{Hung2023}, a voltage-to-time converter is added to amplify and linearize the RC discharge profile of ReRAM cells, improving resolution but introducing power and latency overhead. In both cases, the memory is not directly integrated into the delay path, limiting efficiency, scalability, and reconfigurability.

To address these challenges, other works employ NVMs outside the signal path~\cite{Saha2024,Yin2024}, using their stored states to control switches rather than modulating delay directly. In this scheme, the memory content is read and used to digitally or analogously control the delay of a DE. Although this avoids issues related to high ON resistance in the signal path, it breaks the in-memory computing paradigm by introducing a separate read-and-control stage. Furthermore, it adds routing and circuit complexity, increasing area and decreasing the efficiency. 

\begin{figure}[!t]
    \centering
    \includegraphics[width=0.9\columnwidth,trim={0.6cm 0.6cm 0.6cm 0.6cm},clip]{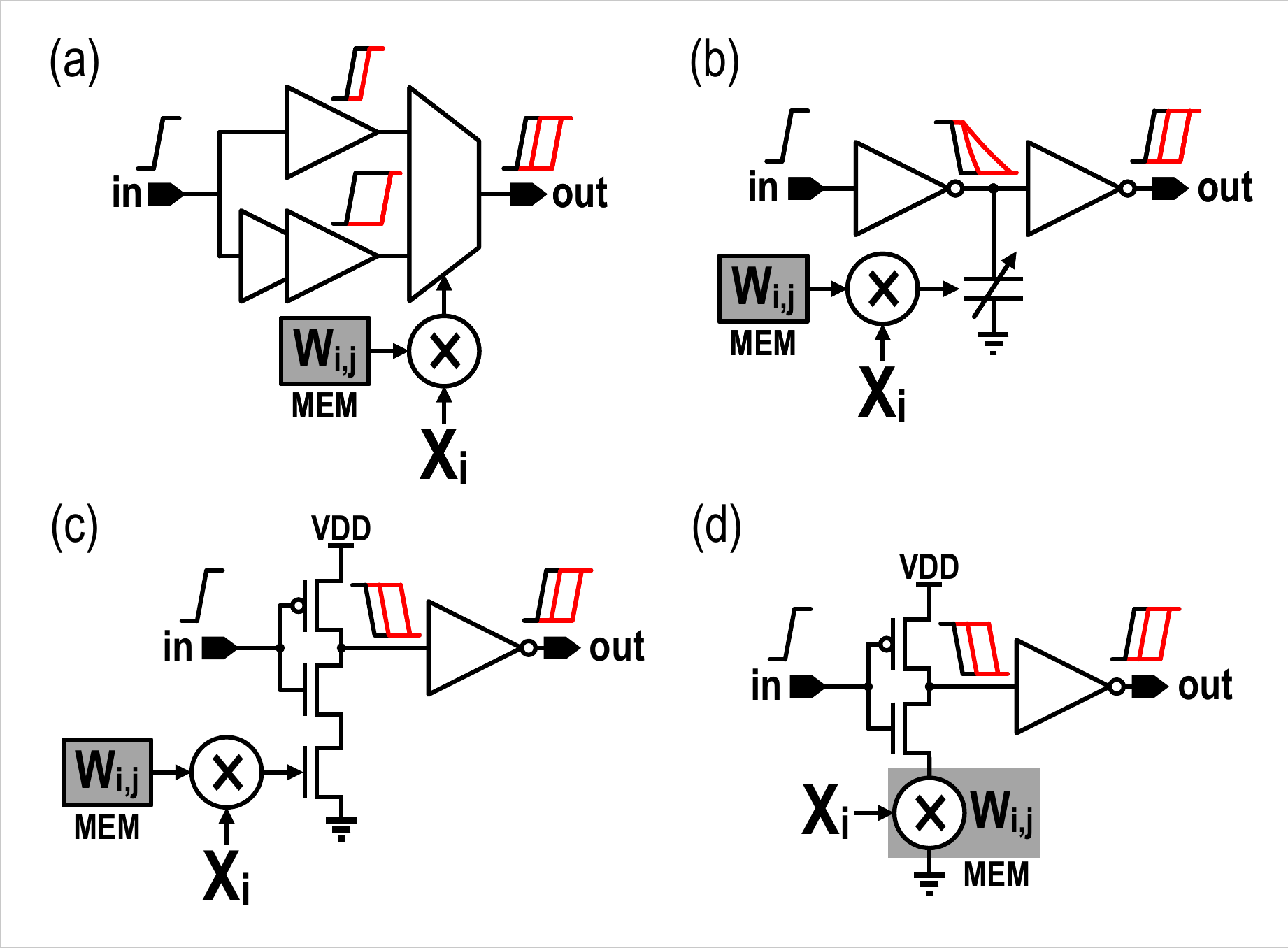}
    \caption{Representative DE topologies integrated with memory (MEM) elements, where delay encodes the weight–activation multiplication result. (a) Multiplexer-based path selection. (b) Capacitive load modulation. (c) Tail current gating via external memory readout. (d) Programmable tail device for direct delay control..}
    \label{fig:de_types}
\end{figure}

To provide a broader perspective on existing time-domain multiplication schemes, Fig.~\ref{fig:de_types} illustrates several representative DE implementations. These topologies differ in how $W_{i,j}$ and $X_i$ interact to modulate $t_d$. In \rev{Fig.~\ref{fig:de_types}(a)}, $W_{i,j}$ is stored in a memory element, and $X_i$ is used to evaluate the multiplication result, which then selects one of two delay paths via a multiplexer; the effective delay reflects the outcome of this time-domain multiply operation. This structure, demonstrated using an OAI-based SR-latch in\cite{Lou2024}, does not use nonvolatile memory, but the general concept can be extended to NVM-based storage. While simple, this approach requires multiplexers, adding area and limiting density.

In \rev{Fig.~\ref{fig:de_types}(b)}, $t_d$ is modulated by enabling or disabling capacitive loads between inverter stages, controlled by $X_i$-$W_{i,j}$ combinations. Yin \textit{et al}.\cite{Yin2024} implemented this topology using FeFETs to demonstrate both XOR- and AND-based MAC operations. The scheme supports reconfigurability by adjusting logic mappings at the input, but it relies on external analog switches and capacitors, limiting integration and achieving large $\Delta_s>1~\mu s$ due to the use of discrete components on a breadboard. An externally controlled CSI topology is illustrated in \rev{Fig.~\ref{fig:de_types}(c)}, where the tail current source is gated by the multiplication result. In\cite{Saha2024}, an STT-MRAM cell stores $W_{i,j}$, and its readout voltage---corresponding to the product of the stored weight and the applied activation---is used to control the tail NMOS transistor of the CSI, modulating $t_d$. A similar approach is also used in SRAM-based designs, such as~\cite{Song2021}, where the readout of a memory cell encodes the multiplication result and enables or disables the current path. This method relies on peripheral readout circuits and additional control logic, increasing system overhead.

Finally, a DE where the discharge current is modulated by a programmable tail transistor is shown in \rev{Fig.~\ref{fig:de_types}(d)}. This DE enables direct control of $t_d$. A CSI with a Fe-FinFET tail is proposed in~\cite{Luo2021} to realize this concept. However, the demonstrated system includes only a few delay stages with delay steps ($\Delta_s$) larger than 10~$\mu$s and lacks full on-chip integration, such as data converters, calibration circuits, and peripheral logic. Furthermore, the number of programmable delay states is limited.

\section{FeFET TD-nvIMC Macro Design and Operation}
\label{sec:arch}

\begin{figure}[!t]
    \centering
    \includegraphics[width=\columnwidth,trim={0.6cm 0.6cm 0.6cm 0.6cm},clip]{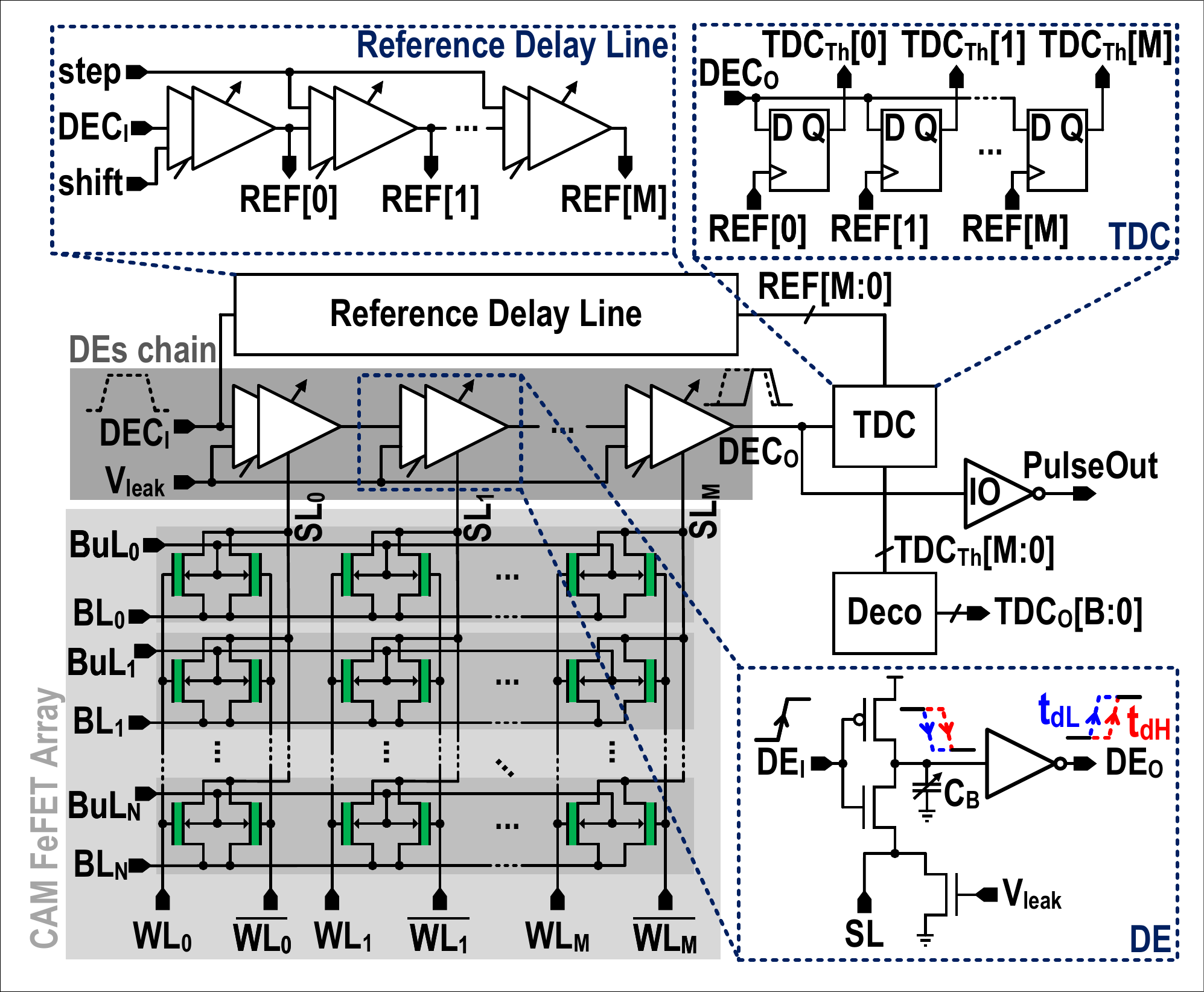}
    \caption{Proposed TD-nvIMC architecture composed of a DE chain with CSI whose tail is implemented by a CAM cell and a leaker driving an inverter, TDC with tunable reference delay line, and I/O for observability. }
    \label{fig:arch}
\end{figure}

The proposed macro, implemented in GlobalFoundries 28SLPe~\cite{Trentzsch2016},   consists of a CAM FeFET array implemented using the C-AND topology~\cite{Dahan2022}, a DE chain, TDC, a reference delay line, a high-speed I/O driver, decoders, and test circuitry, as shown in Fig.~\ref{fig:arch}. The FeFET cells have width ($W_{FeFET}$) and length ($L_{FeFET}$) of 90~nm. In this section, the memory array implementation is presented together with write-disturb prevention and MLS scheme using the bulk. Then, the DE and CSI design and operation are described. Finally, the TDC implementation is discussed. 

\subsection{Memory Array Design, Write-Disturb Prevention}
\label{sec:memory}
\begin{figure}[!t]
    \centering
    \includegraphics[width=\columnwidth,trim={0.6cm 0.6cm 0.5cm 0.6cm},clip]{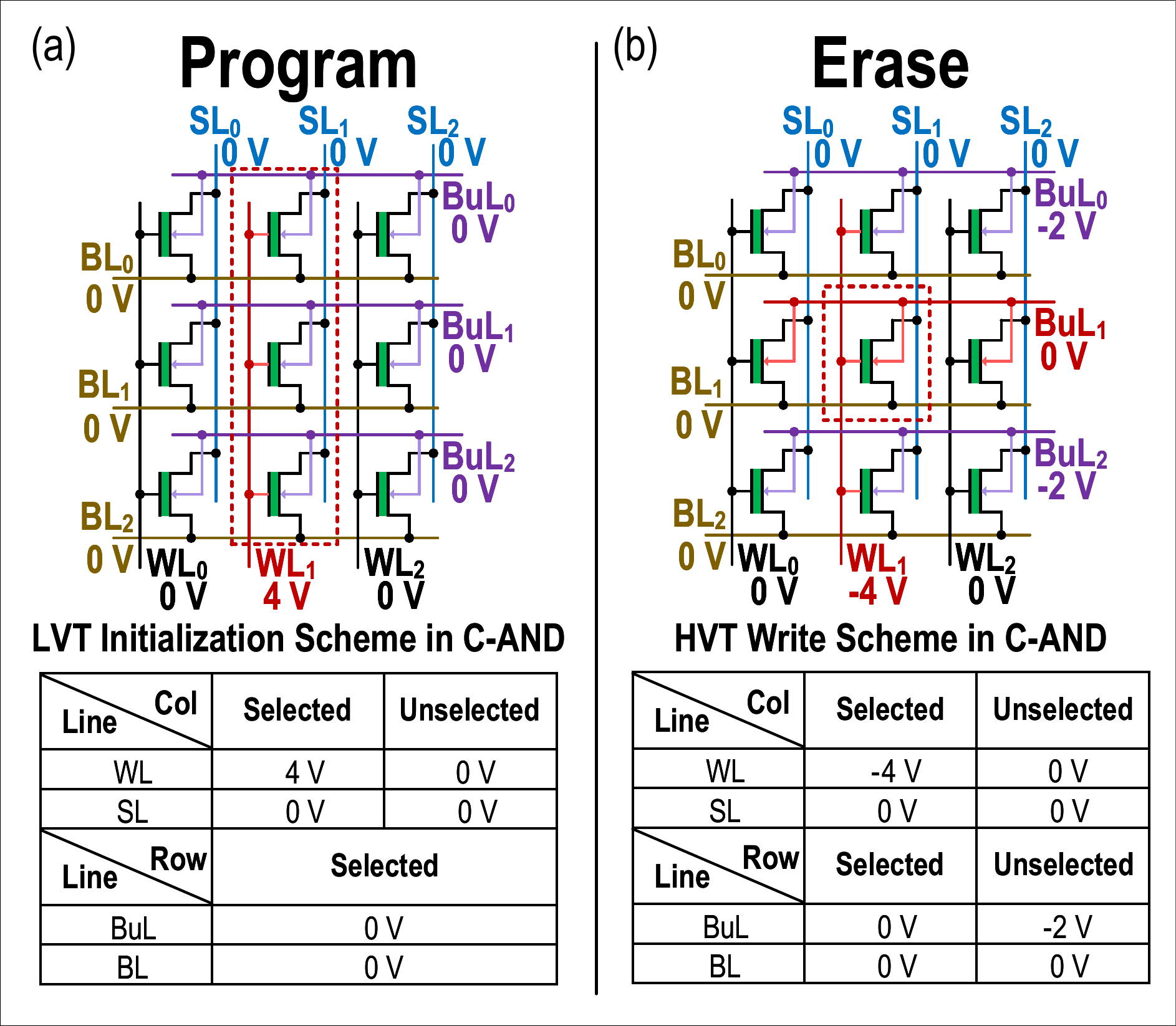}
    \caption{C-AND schematic and write scheme. \rev{(a)} First, all cells in the selected column are programmed to LVT. \rev{(b)} Erase is performed by selecting the target cell ($WL=-4~V$) and setting $BuL=0~V$. Write-disturb is prevented on adjacent cells in the column with $BuL=-2~V$.}
    \label{fig:cand}
\end{figure}

The CAM cell consists of two FeFETs storing complementary values. The CAM follows a C-AND array architecture~\cite{Dahan2022}, where select lines (SLs) and word lines (WLs) are shared column-wise, while bit lines (BLs) and bulk lines (BuLs) are shared row-wise (see Fig.~\ref{fig:cand}). Furthermore, each row is implemented in an independent bulk using a triple-well process.
Programming to LVT is achieved by applying 4~V to the selected WL, while keeping SL, BL, and BuL at 0~V. At the array level, isolating individual rows during programming is not possible by raising BuL. As a result, the entire column is programmed simultaneously.

\begin{figure}[!t]
    \centering
    \includegraphics[width=\columnwidth,trim={0.6cm 0.6cm 0.5cm 0.6cm},clip]{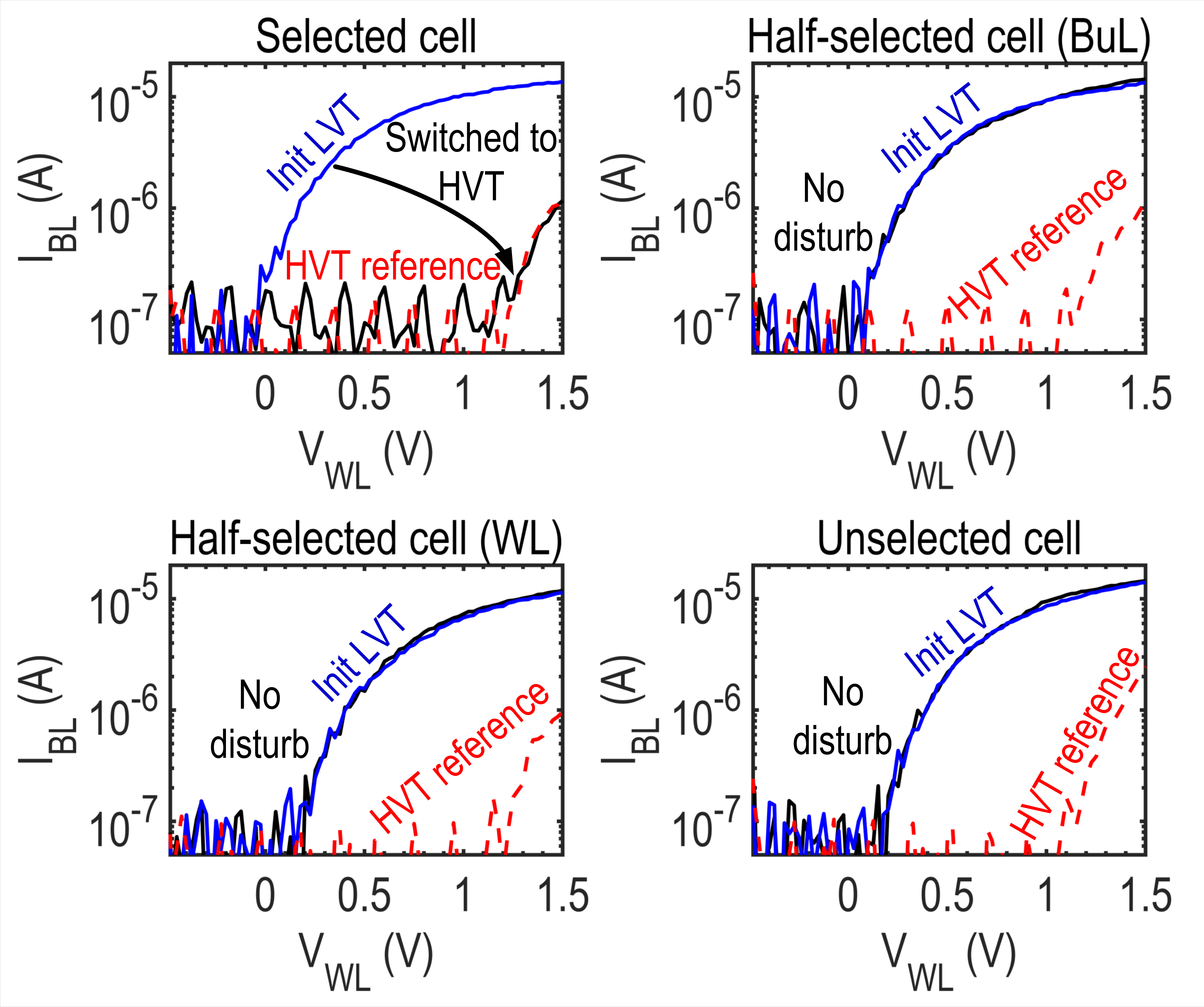}
    %\caption{\rev{Array-level write-disturb prevention using bulk biasing. Measured $I_{BL}$--$V_{WL}$ characteristics for a selected cell, half-selected cells sharing the same BuL or WL, and an unselected cell. Black curves indicate the final state, showing that only the selected cell switches to HVT after erase, while all others show no disturb.}}
    \caption{\rev{Array-level write-disturb prevention using bulk biasing. Measured $I_{BL}$--$V_{WL}$ characteristics for a selected cell, half-selected cells sharing the same BuL or WL, and an unselected cell. Only the selected cell switches to HVT after erase, while all other cases show no disturb.}}
    \label{fig:cand_meas}
\end{figure}

To erase a cell to HVT, -4~V is applied to the WL of the selected cell, while the BL, SL, and BuL are held at 0~V. To prevent write disturbance, the other cells in the same column are isolated by biasing the BuL to -2~V. This negative bulk bias applied to unselected rows suppresses unintended polarization switching during write pulses for the cells in the same column. Triple-well isolation ensures that each row has a dedicated bulk node, enabling independent control of the BuL signals per row. Jiang \textit{et al}.~\cite{Jiang2022} showed that this scheme is effective at the device level, demonstrating that bulk biasing can prevent switching in individual FeFETs. In this work, we validate this approach at the array level through experimental measurements (see \rev{Fig.~\ref{fig:cand_meas}}), confirming that only the selected cell switches to HVT while all unselected cells remain undisturbed. To the best of our knowledge, this is the first array-level validation of bulk-assisted write-disturb prevention in FeFET-based C-AND arrays.
% To erase a cell to high threshold voltage (HVT), -4~V is applied to the WL of the selected cell, while BL, SL, and BuL are kept at 0~V. The other cells in the same column are isolated by biasing the BuL to -2~V to prevent write-disturb.

\subsection{Multilevel States using Bulk}
\label{sec:MLS}
\begin{figure}[!t]
    \centering
    \begin{subfigure}{1\columnwidth} % Adjust width as needed
        \centering
         \includegraphics[width=0.8\columnwidth,trim={0.6cm 0.6cm 0.6cm 0.6cm},clip]{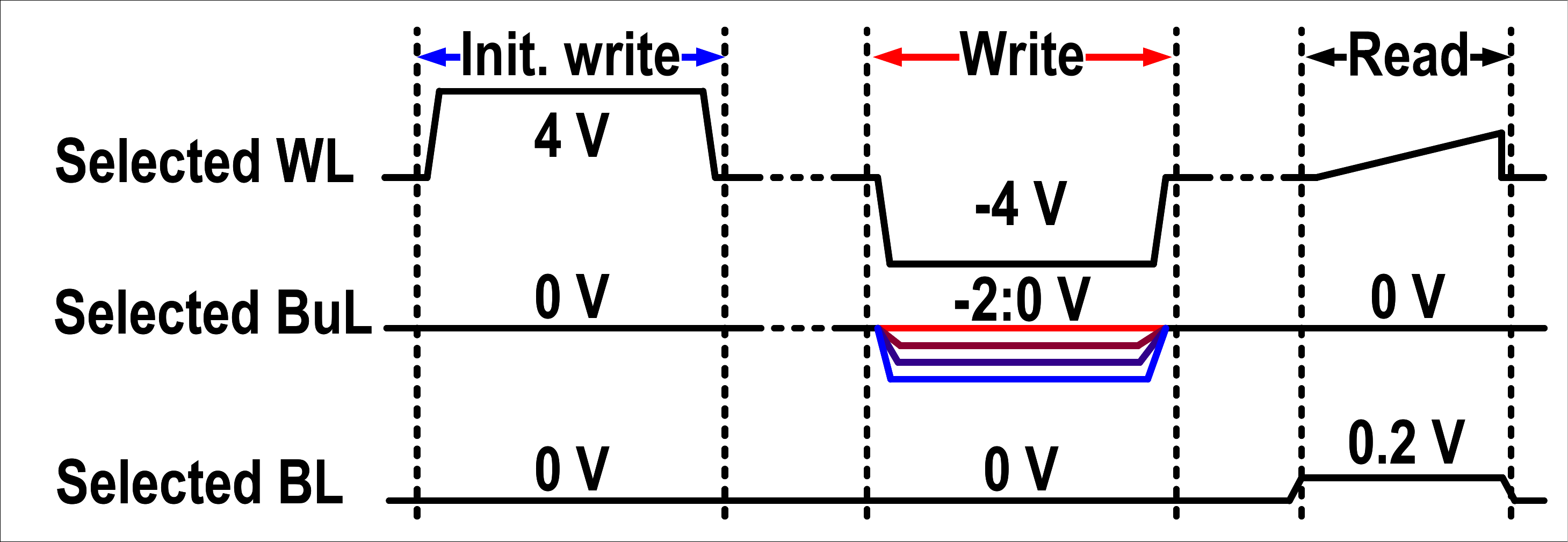}
         \caption{}
    \end{subfigure}
    \begin{subfigure}{0.49\columnwidth} % Adjust width as needed
         \includegraphics[width=\columnwidth,trim={0.6cm 0.6cm 0.6cm 0.6cm},clip]{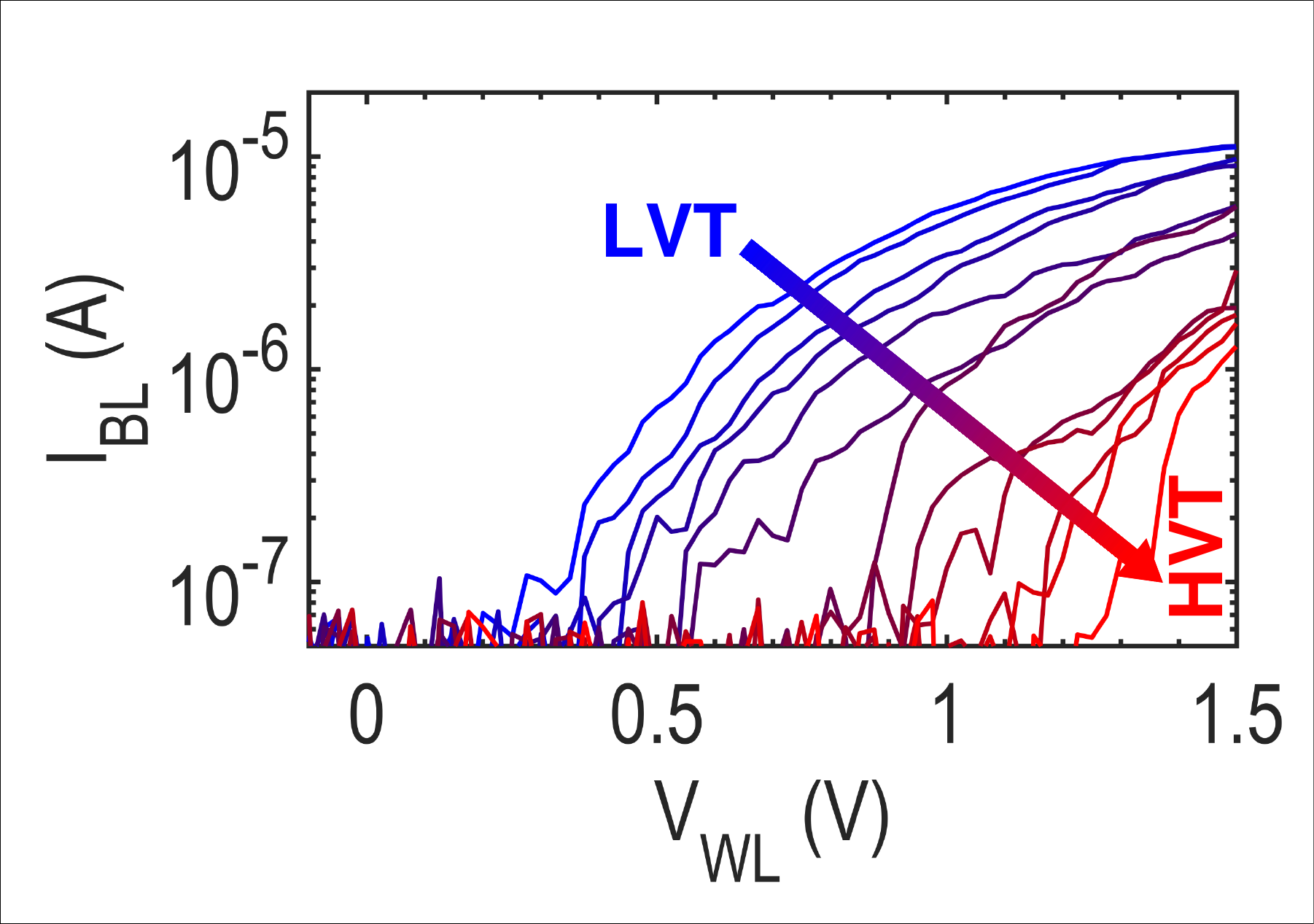}
         \caption{}
    \end{subfigure}
      \begin{subfigure}{0.49\columnwidth} % Adjust width as needed
         \includegraphics[width=\columnwidth,trim={0.6cm 0.6cm 0.6cm 0.6cm},clip]{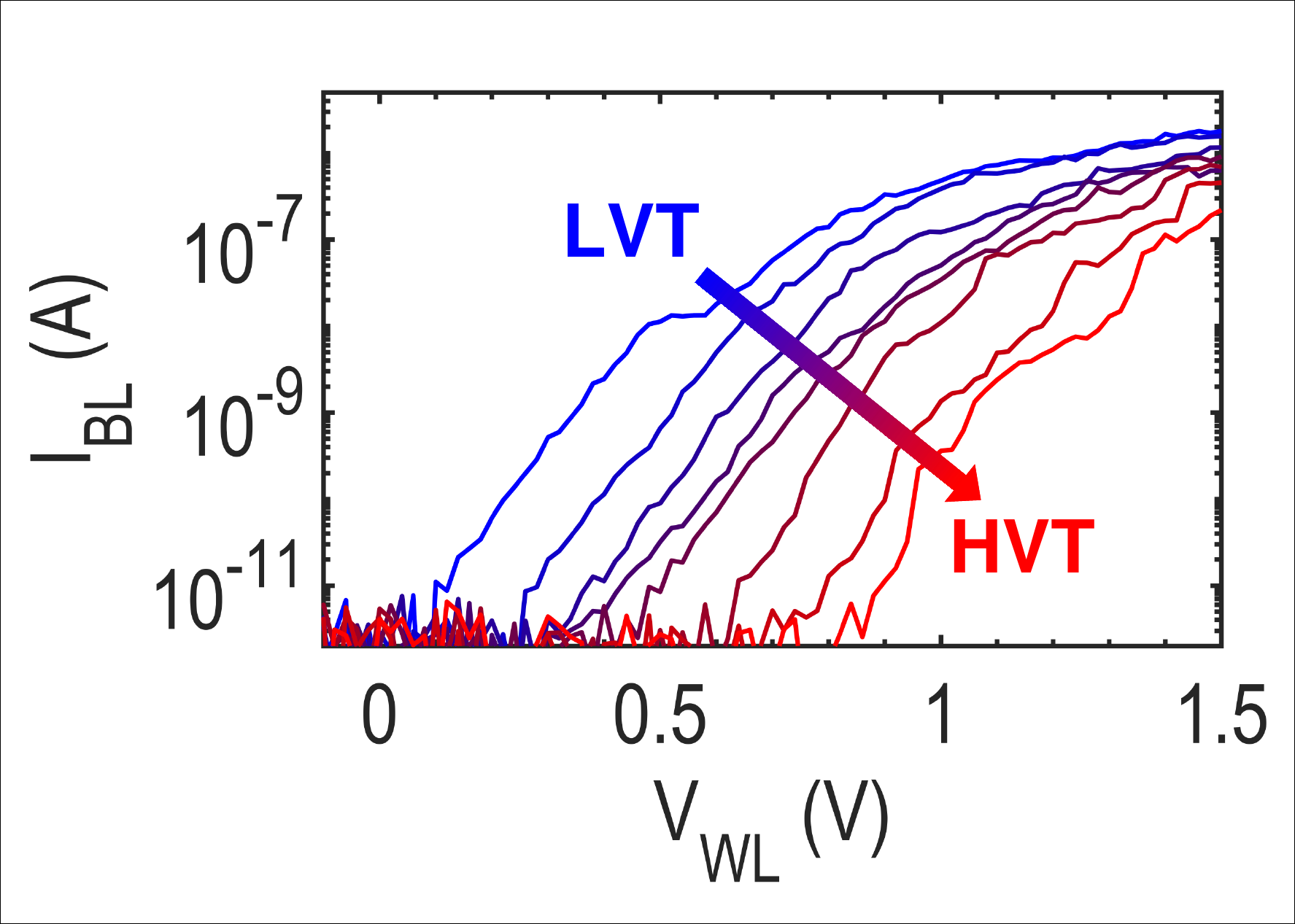}
         \caption{}
    \end{subfigure}
    \caption{ MLS  using bulk biasing. (a) WL, BuL, and BL waveform diagram illustrating the sequence of LVT initialization, partial erase for MLS, and read with (b) fast and (c) slow WL voltage sweep. }
    \label{fig:mls_IVmeas}
\end{figure}
We introduce a new scheme for generating MLS within the FeFET array using partial erase via BuL control. Starting from a fully programmed LVT state, MLS is achieved by fixing the WL to –4~V and sweeping the BuL between –2~V and 0~V [see Fig.~\ref{fig:mls_IVmeas}(a)]. This partial erase mechanism incrementally increases the $V_{T}$ of the selected cell, enabling access to intermediate states. Unlike partial programming, which poses a higher risk of disturbance due to positive write pulses, the proposed approach takes advantage of the disturb-robust erase mechanism in combination with triple-well isolation to ensure safe tuning of $V_{T}$ without affecting adjacent cells.

The behavior is verified using WL voltage sweeps at two different rates during readout: a fast linear ramp and a slow ramp, as shown in Fig.~\ref{fig:mls_IVmeas}(b-c). These results confirm the feasibility of finely tuned, robust multilevel storage using a single-ended voltage bias scheme, without requiring modifications to the FeFET structure or peripheral circuits. To the best of our knowledge, this is the first demonstration of MLS generation using bulk-assisted partial erase in a FeFET array. This capability is crucial for calibration and fine-grained tuning in analog and time-domain applications.

\subsection{TD-nvIMC Macro Operation}
Each row in the CAM array stores a weight vector, with each CAM cell holding $W_{i,j}$ and its complement ($\overline{W_{i,j}}$). During initialization, the array is globally programmed to LVT, followed by column-wise selective erase to store $W_{i,j}=0$ or $W_{i,j}=-1$, for AND and XOR MAC, respectively, using the MLS scheme discussed in Section ~\ref{sec:MLS}. Operations are carried out row-by-row: a single row is activated by grounding its BL, while all other BLs remain floating (Hi-Z), ensuring isolated participation of that row in the computation.
Activations $X_i$ are applied to the WL and $\overline{\text{WL}}$ lines depending on the MAC configuration (namely, XOR or AND). 

The input pulse is injected into the DE chain, where each stage incrementally modulates its $t_d$ based on the local $W_{i,j}$–$X_i$ pair. The $T_D$ across the chain effectively encodes the result of the MAC operation in the time domain. This delayed output pulse is digitized by a TDC, which consists of a reference delay line (RDL) and a bank of flip-flops (FFs) for sampling. The TDC is configured via programmable \textit{step} and \textit{shift} control signals to adapt its dynamic range and resolution to the application requirements. The resulting thermometer-coded output is then converted to binary and made available off-chip or visualized directly via the I/O driver. %Each subarray operates autonomously, incorporating its own CAM, DE chain, and TDC. This self-contained modular architecture enables parallelism and scalability while avoiding global synchronization bottlenecks and minimizing interconnect complexity.

\subsection{DE and CSI}
Each SL is connected to a DE, composed of a CSI, loaded by a capacitor bank ($C_B$) and an inverter stage to restore the pulse polarity and sharpen the edges (see Fig.~\ref{fig:arch}). The CSI is implemented by an inverter whose tail is connected to a FeFET CAM cell by its SL and a parallel NMOS leaker. The CAM cell provides a conditional discharge path whose resistance depends on the combination of the stored $W_{i,j}$ and the applied $X_i$. The effective resistance of the CAM cell, $R_{CAM}$, is the result of the two FeFETs in the cell and is given by:
\begin{dmath}
    R_{CAM}=\frac{L_{FeFET}/W_{FeFET}}{k_{FeFET}(V_{WL}-V_{T,FeFET})}\parallel \frac{L_{FeFET}/W_{FeFET}}{k_{FeFET}(V_{\overline{WL}}-V_{T,\overline{FeFET}})}\rev{,}
    \label{eq:rcam}
\end{dmath}
where $k_{FeFET}$ is the transconductance factor, and $V_{T,FeFET}$ and $V_{T,\overline{FeFET}}$ are the threshold voltages of the main and complementary FeFETs, respectively. $V_{WL}$ and $V_{\overline{WL}}$ are the gate voltages applied to the selected WL, determined by $X_i$. 

The leaker provides a controlled discharge path when both FeFETs are effectively off, ensuring a well-defined slow delay state ($t_{dH}$). Its resistance is defined by:
\begin{equation}
    R_{leaker}=\frac{L/W}{k(V_{leak}-V_T)}.
    \label{eq:rleaker}
\end{equation}
Here, $L$, $W$, $k$, and $V_T$ represent the transistor length, width, transconductance, and threshold voltage, respectively. The total effective resistance, $R_{eff}$, is the parallel combination of $R_{CAM}$ and $R_{leaker}$, in series with the pull-down NMOS of the CSI:
\begin{equation}
    R_{eff}=R_{CAM}\parallel R_{leaker} + R_{NMOS}.
    \label{eq:reff}
\end{equation}
The CSI fall delay, $t_{df,CSI}$, is proportional to the product of this $R_{eff}$ and the capacitor bank $C_B$:
\begin{equation}
    t_{d,CSI}\propto R_{eff}C_B~,
    \label{eq:tdcsi}
\end{equation}
Thus, the leaker voltage ($V_{leak}$) sets the high delay ($t_{dH}$), while the low delay ($t_{dL}$) is set by the threshold voltages of both FeFETs in the CAM cell and the applied gate voltages $V_{WL}$ and $\overline{V_{WL}}$. The capacitor bank is also used to calibrate the delay step $\Delta s$ via programmable loading.

\subsection{TDC}
The output of the DE chain (DEC\textsubscript{O}) is sampled by a TDC, which generates a digital code proportional to $T_D$. A Flash-type architecture is adopted to allow for fast, low-latency conversion. The TDC consists of a chain of D-type FFs that simultaneously sample DEC\textsubscript{O} against a set of reference signals REF[M:0]. These reference signals are evenly spaced in time and generated by a tunable RDL, which supports adjustable resolution and phase alignment through the \textit{step} and \textit{shift} control inputs.

Each FF samples the DEC\textsubscript{O} pulse relative to its corresponding reference signal REF[i], generated by the calibrated RDL. As a result, the TDC output is initially expressed as a thermometer code TDC\textsubscript{Th}[M:0], which is then converted to a binary value TDC\textsubscript{O}[B:0] before being driven off-chip. The binary output reflects the relative arrival time of the DEC\textsubscript{O} edge with respect to the reference signal array, with earlier edges producing lower binary codes at the TDC output and later edges producing higher ones.

The time resolution of the TDC is defined by the spacing between adjacent reference signals REF[i], and is determined by the delay step of the RDL, controlled through the \textit{step} input. This resolution is configured to match the expected delay difference $\Delta s$ between adjacent MAC results in the DE chain, ensuring that all valid outputs can be reliably distinguished. By aligning the TDC resolution with the discrete delay levels produced by the DE chain, the system avoids quantization overlap and enables clean, one-to-one mapping of delay levels to digital codes.

The Flash architecture enables fully parallel sampling of DEC\textsubscript{O} without the need for high-speed global clocking. The resulting thermometer code encodes how many reference signals have transitioned before the arrival of DEC\textsubscript{O}, effectively quantizing the delay within the reference interval. After conversion to binary, each digital code corresponds to a specific quantized delay level, enabling robust classification of MAC outcomes with sub-nanosecond (sub-ns) precision. 
%To support observability and debugging, DEC\textsubscript{O} is also routed to a 50~$\Omega$ I/O driver with 100~ps edge transitions, allowing direct probing of the analog timing behavior using high-speed oscilloscopes and complementing the digitized output captured by the TDC.

\section{Reconfigurable IMC Logic Operation}
\label{sec:imc}

The proposed macro supports multiple logic modes, enabling both MAC and Boolean operations using the same time-domain infrastructure. Each mode exploits the conditional behavior of the DEs, whose output delay depends on the match between the stored weight $W_{i,j}$ and the applied activation $X_i$. This section details the configuration of the control signals for each supported logic operation and describes how the output delay encodes the result.

\subsection{XOR-MAC}
\begin{figure}[!t]
    \centering
    \includegraphics[width=\columnwidth,trim={0.6cm 0.6cm 0.6cm 0.6cm},clip]{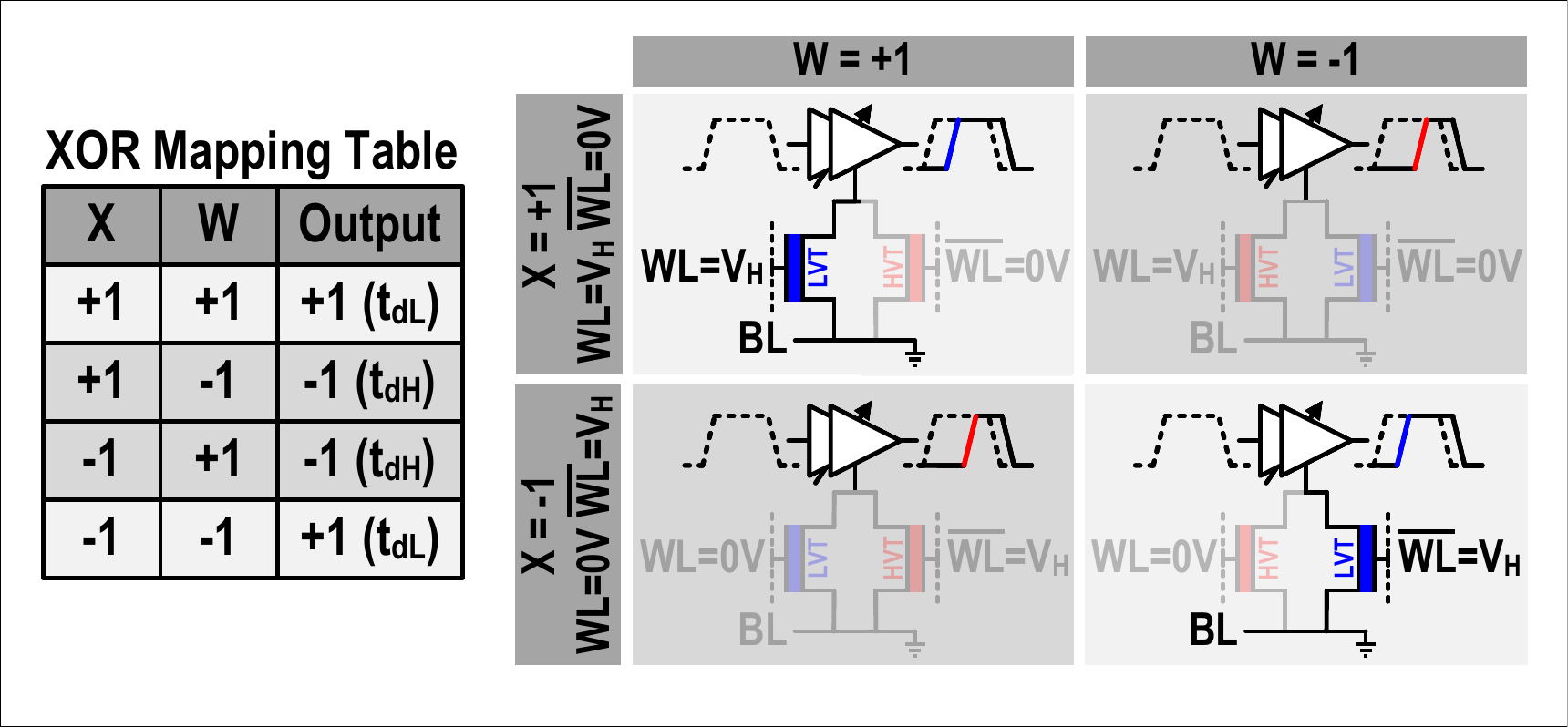}
    \caption{ XOR-based MAC truth table and schematic operation.  }
    \label{fig:mac_xor_op}
\end{figure}

The XOR-MAC operation is performed by grounding the selected BL, while setting all other BLs to Hi-Z (Fig.~\ref{fig:mac_xor_op}). The activation inputs $X_i$ are applied to the WL and their complements ($\overline{X_i}$) to $\overline{\text{WL}}$. Logic `1' is set to $V_H$, which is set during the calibration phase (see Sec.~\ref{sec:calibration}). A match, $X_i=W_{i,j}$, leads to a fast discharge path through a LVT FeFET with a high gate voltage, resulting in $t_{dL}$ delay, while $X_i\neq W_{i,j}$ disables the fast discharge path either becaue the FeFET being in HVT state, or due to the gate voltage being low (0~V). In this case, the DE discharges only through the leaker, producing $t_{dH}$ delay. This conditional delay behavior stems from the effective resistance at the tail of the CSI, modeled as $R_{CAM} \parallel R_{leaker}$, as described by expressions ~\eqref{eq:rcam}--\eqref{eq:reff}.

In the match case, the active LVT FeFET dominates, presenting a low $R_{FeFET,on}$ while the off-path (HVT or low gate voltage or both) remains in the high-resistance state $R_{FeFET,off}$. As a result, the overall resistance $R_{FeFET,on} \parallel R_{FeFET,off} \parallel R_{leaker}$ simplifies approximately to $R_{FeFET,on}$, enabling fast discharge. In contrast, for mismatched inputs, both FeFETs are effectively off and the leaker provides the only conductive path. Here, the effective resistance becomes the parallel combination of two high-resistance FeFETs and the leaker, \textit{i.e.}, $R_{FeFET,off} \parallel R_{FeFET,off} \parallel R_{leaker}$ which is orders of magnitude higher than in the match case. \rev{Thus}, implementing the XOR logic function. The delay is then accumulated in the DE chain to implement the final MAC operation. 

\subsection{AND-MAC}
\begin{figure}[!t]
    \centering
    \includegraphics[width=\columnwidth,trim={0.6cm 0.6cm 0.6cm 0.6cm},clip]{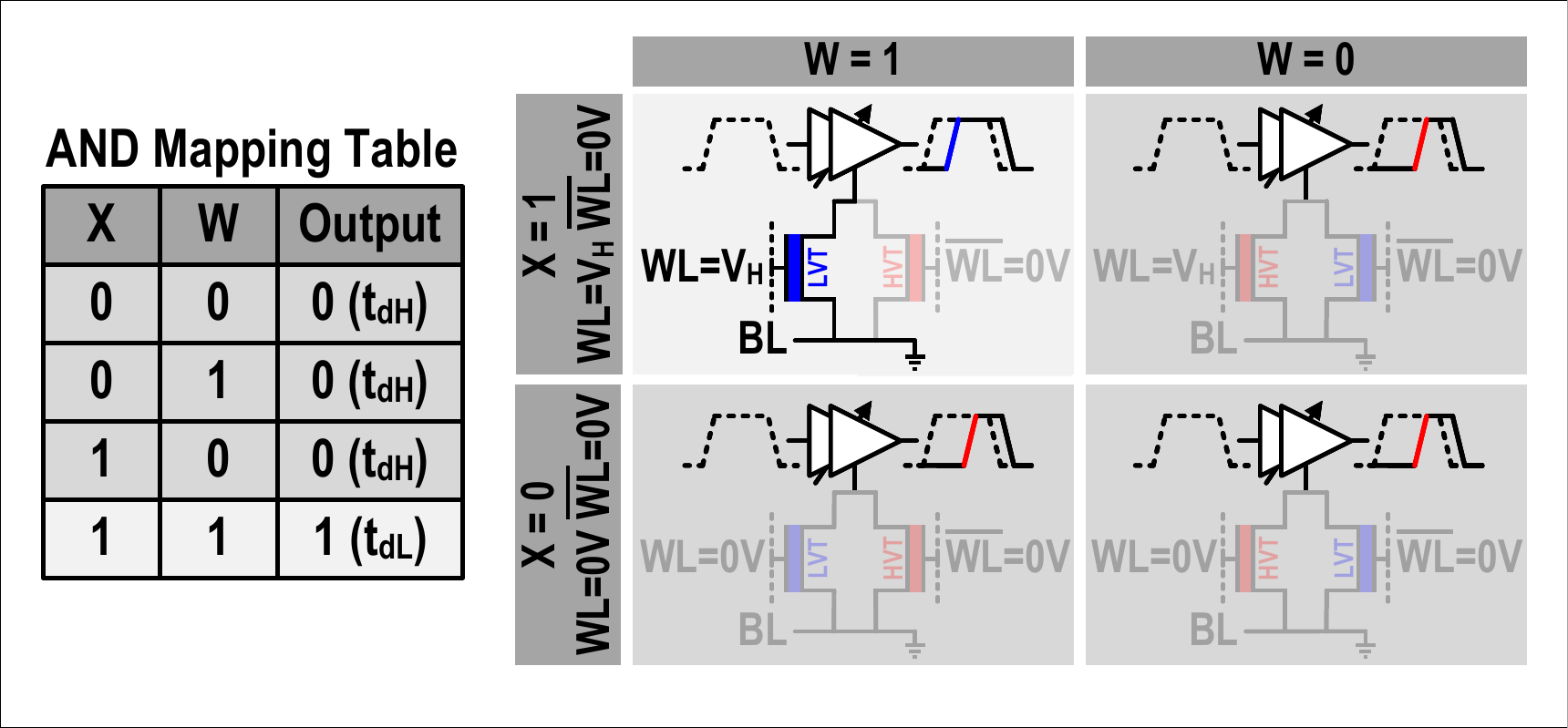}
    \caption{ AND-based MAC truth table and schematic operation.  }
    \label{fig:mac_and_op}
\end{figure}

The AND-MAC operation is performed by grounding the selected BL, while setting all other BLs to Hi-Z, as shown in Fig.~\ref{fig:mac_and_op}). In this configuration, $\overline{\text{WL}}$ is always grounded, ensuring that the complementary FeFET remains off and does not contribute to the discharge path. Fast discharge occurs only when the main FeFET is in the LVT state (logic `1') and the corresponding $X_i$ is also `1', applied as a high gate voltage on the WL. This results in a low-resistance path through the active FeFET, producing the short delay $t_{dL}$. In all other cases, whether the FeFET is in HVT ($W_{i,j}=0$), or $X_i = 0$ (low gate voltage), the discharge path through the CAM cell remains blocked. In such cases, the DE discharges only through the leaker, resulting in a longer delay $t_{dH}$. The delay is then accumulated across the DE chain to implement the AND-based MAC operation.

\subsection{Boolean Logic}
In addition to MAC operations, the reconfigurable macro supports in-memory logic operations such as OR and AND, as well as full adder operations, performed directly on the stored FeFET states. These operations are executed per row, using a subset of $k$ selected columns out of total $M$. A row is selected by grounding the corresponding BL, while all other BLs are set to Hi-Z, as in the MAC mode. The participating columns are selected by applying $V_H$ to their WLs, while grounding all other WLs and all \rev{$\overline{\text{WL}}$} signals (including for selected cells). This ensures that the DEs of the unselected cells remain in their high-delay state, unaffected by the stored memory (FeFET state).

For in-memory logic AND, the output distinguishes the case where all $k$ selected cells store `1'. Only in this case will the DEs in all $k$ columns exhibit $t_{dL}$ delay, while in the remaining $(M-k)$ columns (DEs) contribute $t_{dH}$ delay. The $T_D$ for this case is $t_{dH} \cdot (M-k)+t_{dL} \cdot k$. The output is taken from the REF[M-k] stage of the TDC, which separates this state from other combinations.

For in-memory logic OR, the output identifies the case where all selected $k$ cells store a `0'. Since `0' corresponds to the HVT state, all selected cells in this condition remain off, and their DEs contribute only via the leaker path, resulting in a cumulative delay of $t_{dH} \cdot M$. The output is taken from the last stage of the TDC (REF[M]). Any deviation from this all-zero case (\textit{i.e.}, at least one cell storing logic `1') will lead to at least one DE switching to its faster $t_{dL}$ path, reducing the overall delay and shifting it away from the REF[$M$] point. Thus, logic `1' is decoded on the TDC output when the delay is shorter than the all-zero baseline.

The full adder is implemented similarly to the AND-based MAC operation, with the key difference being that WLs for unselected cells are grounded to exclude them from the computation. This reuse of the memory structure enables seamless integration of arithmetic and logic operations directly within the memory array. The flexibility of the proposed architecture allows the same FeFET-based cells to perform both computational and storage functions efficiently, reducing data movement and supporting in-memory execution of key digital functions.

\begin{figure}[!t]
    \centering
    \includegraphics[width=\columnwidth,trim={0.6cm 0.6cm 0.6cm 0.6cm},clip]{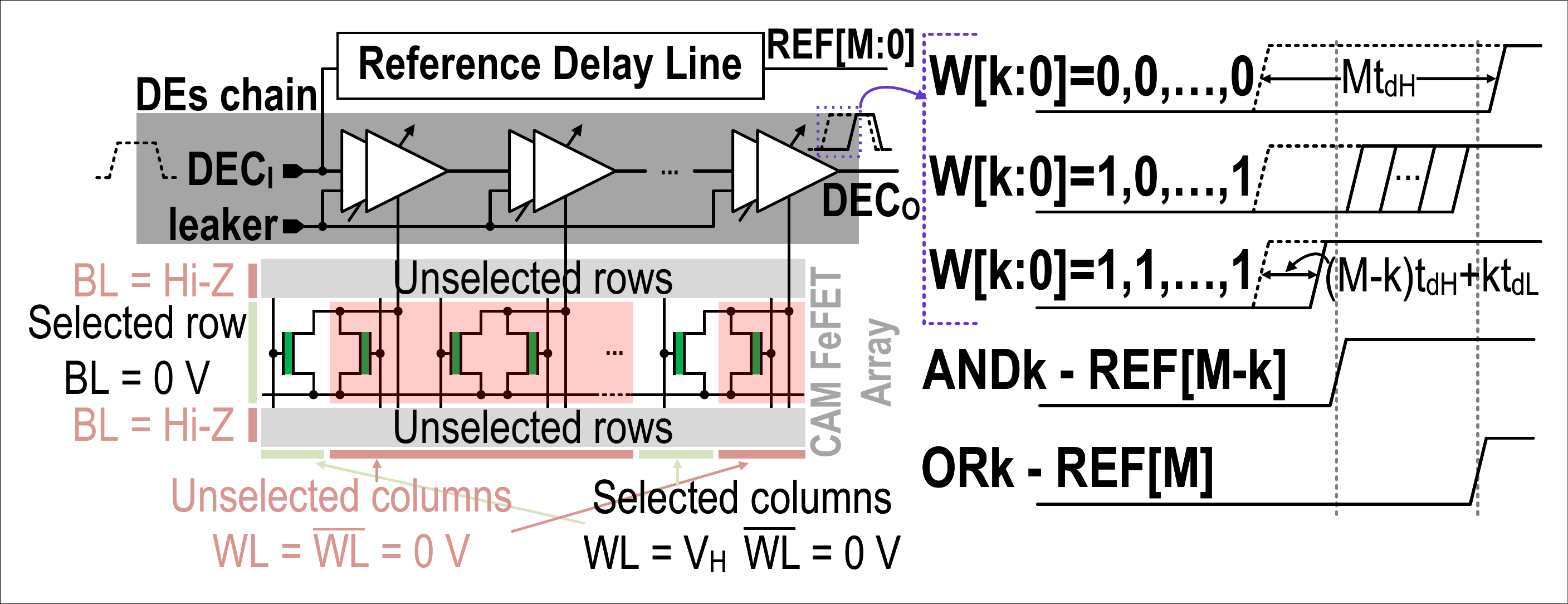}
    \caption{ IMC boolean logic schematic operation and waveform diagram.  }
    \label{fig:logic_sch}
\end{figure}

\vspace{-0.5cm}
\section{Calibration Scheme}
\label{sec:calibration}
To ensure consistent and precise delay modulation across the DE chain, calibration is required to compensate for D2D and DE mismatches. Without calibration, variations in the LVT levels would lead to mismatches in $t_{dL}$, degrading MAC accuracy and time resolution, while affecting the overall energy efficiency. Calibration is performed during weight programming by leveraging MLS and adjusting the BuL bias of individual FeFET cells as detailed in Section~\ref{sec:MLS}.

\begin{figure}[!t]
    \centering
    \begin{subfigure}{1\columnwidth} % Adjust width as needed
        \centering
         \includegraphics[width=\columnwidth,trim={0.6cm 0.6cm 0.6cm 0.6cm},clip]{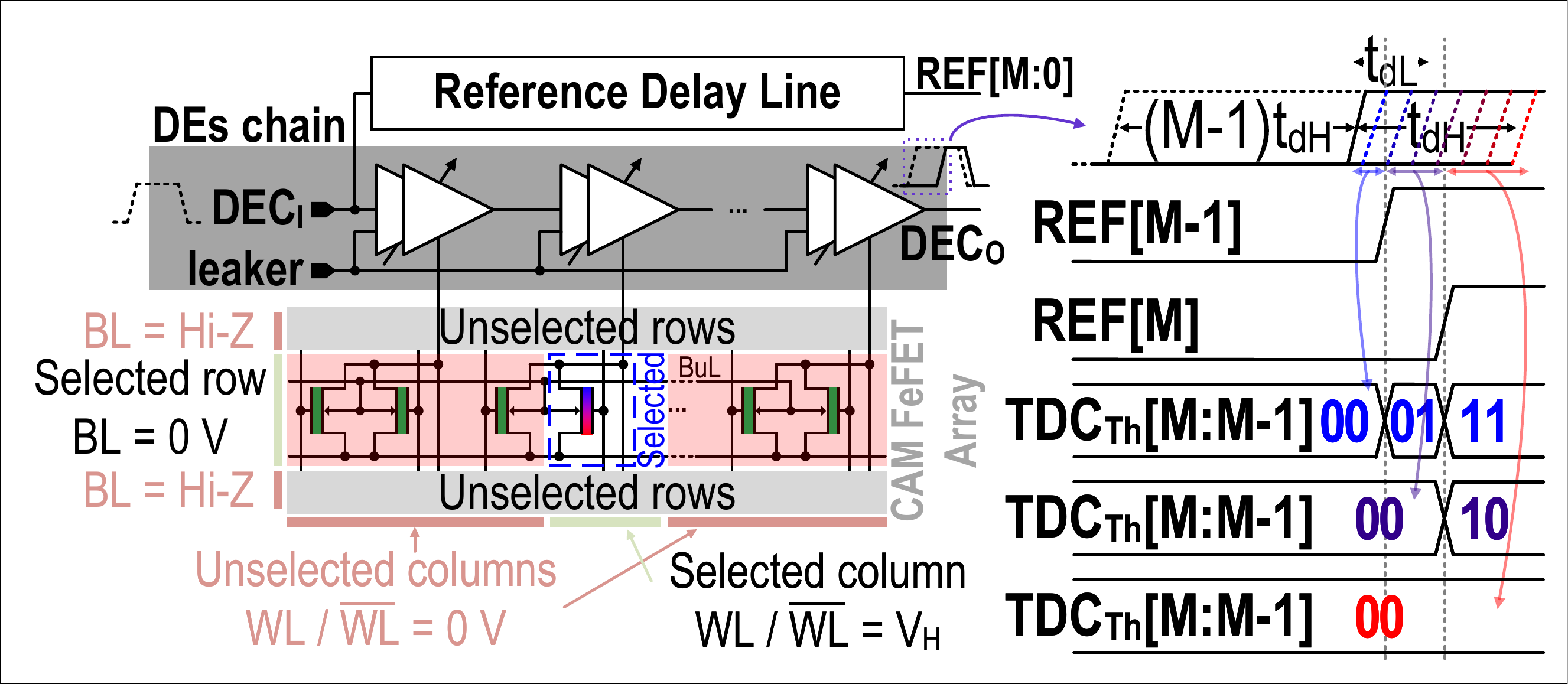}
         \caption{}
    \end{subfigure}
    \begin{subfigure}{0.49\columnwidth} % Adjust width as needed
         \includegraphics[width=1.1\columnwidth,trim={0.6cm 0.6cm 0.6cm 0.6cm},clip]{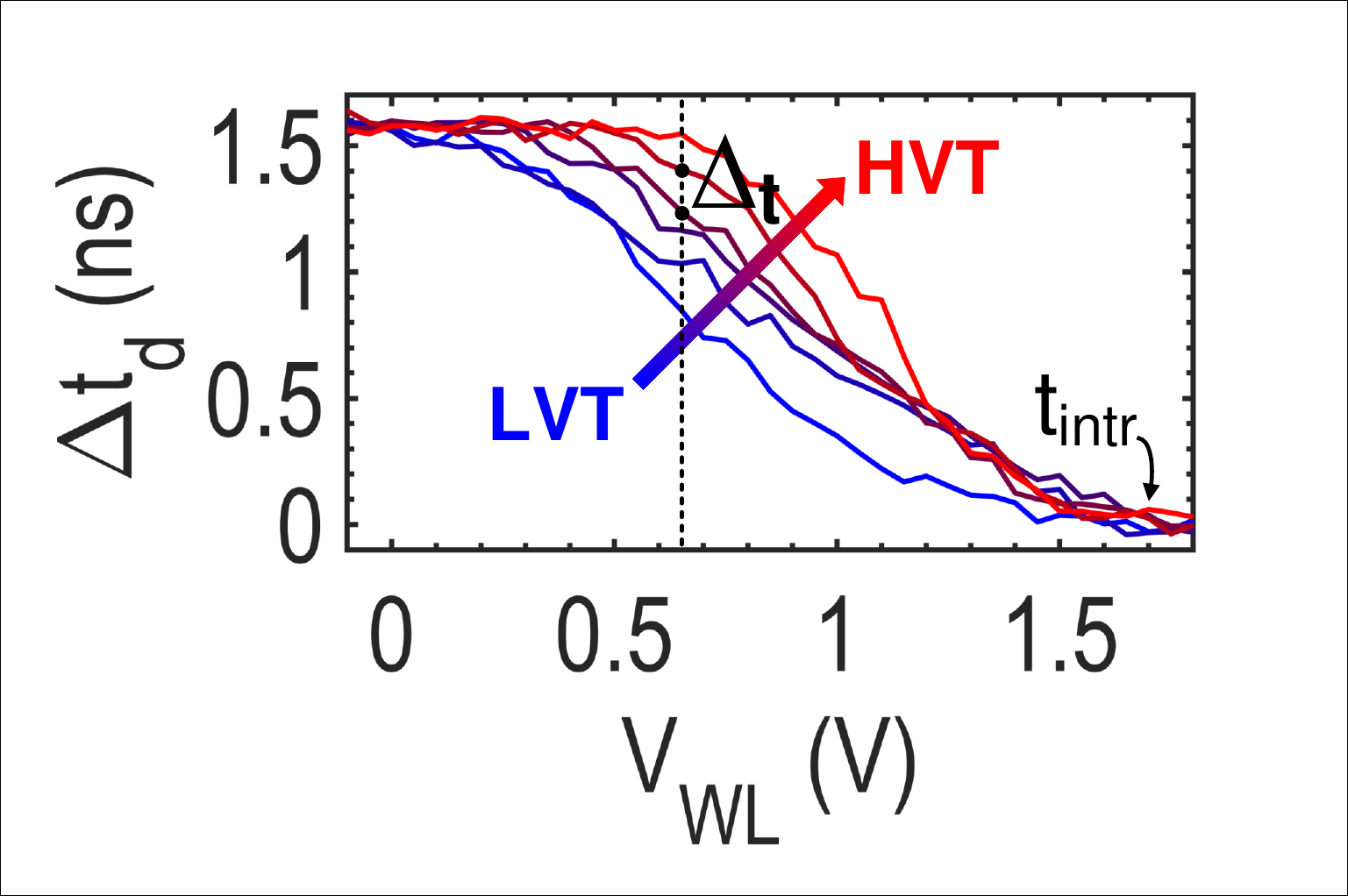}
         \caption{}
    \end{subfigure}
      \begin{subfigure}{0.49\columnwidth} % Adjust width as needed
         \includegraphics[width=1.1\columnwidth,trim={0.6cm 0.6cm 0.6cm 0.6cm},clip]{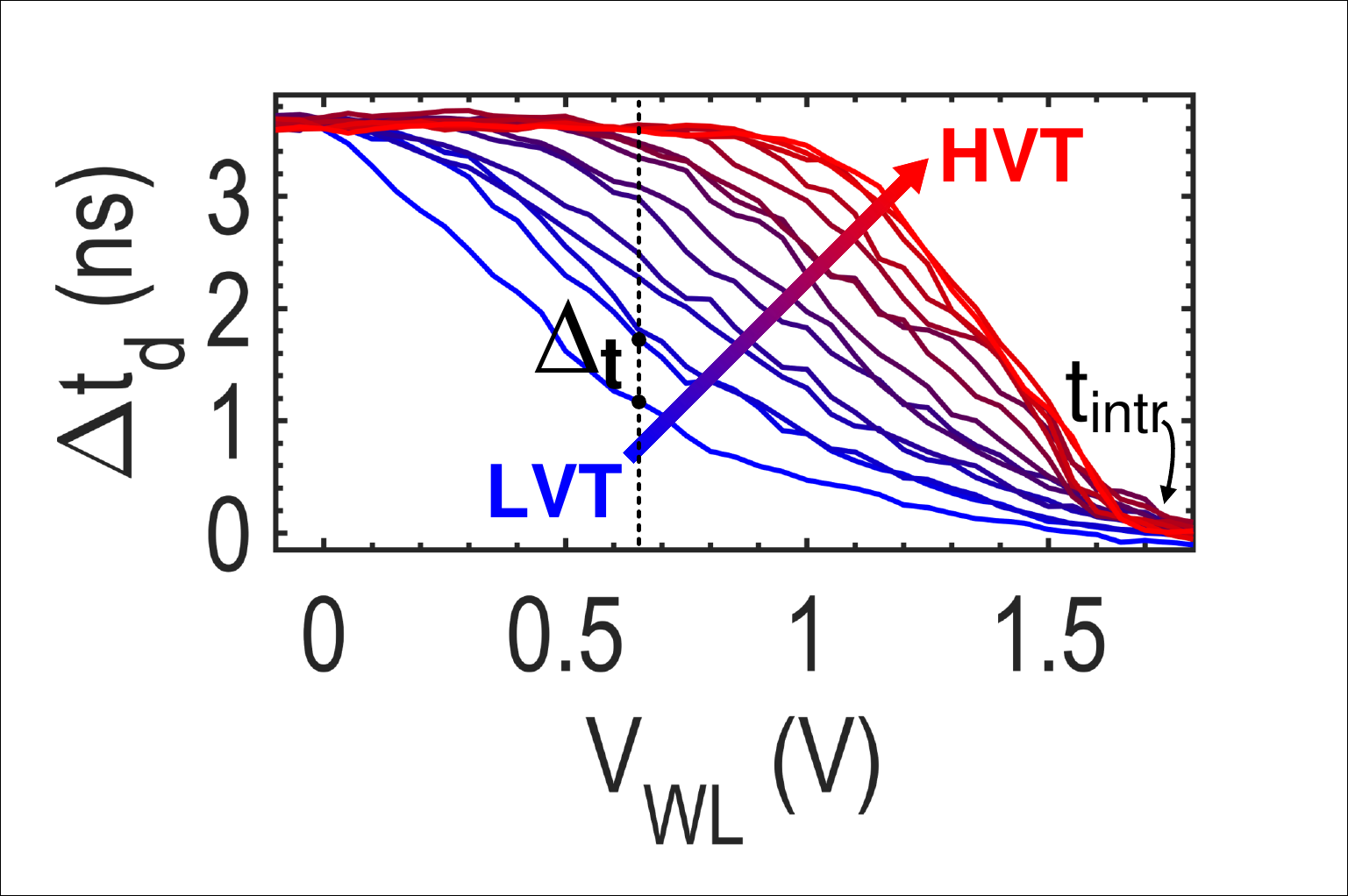}
         \caption{}
    \end{subfigure}
    \caption{ Fixing FeFET and DE mismatch using MLS. (a) Schematic diagram and waveforms of the calibration process. The $V_{T}$ of the selected FeFET is changed until $t_{dL}$ falls within the desired range. Measured $\Delta t_{d}$ for varying $V_{T}$ and WL voltage ($V_{WL}$) for (b) fast and (c) slow silicon dies. A $\Delta t \leq 100~\text{ps}$ is obtained at 0.65~V. }
    \label{fig:mls_calib}
\end{figure}

As illustrated in Fig.~\ref{fig:mls_calib}(a), the target cell is selected by biasing $BL=0~V$ and $WL=V_H$, while all unselected cells are isolated by holding their rows at Hi-Z and columns at 0~V. This configuration ensures that non-targeted DEs contribute a uniform delay ($t_{dH}$), set by the leaker NMOS. 
The TDC monitors the resulting delay, and the BuL voltage is swept from –2~V to 0~V to incrementally adjust the threshold voltage. This process continues until the short-delay condition $t_{dL}$ falls within the target window, enabling tuning with temporal resolution $\leq$100~ps. In the example shown in \rev{Fig.~\ref{fig:mls_calib}(a)}, $t_{dL}$ is tuned until the TDC output transitions to the thermometer code `10' on TDC\textsubscript{Th}[M:M–1].

This calibration is performed each time the FeFET is programmed to the LVT state, enabling fine-grained tuning of $t_{dL}$, associated with logic ‘1’. To evaluate the achievable modulation window, the delay difference ($\Delta t_d=t_d-t_{intr}$, where $t_{intr}$ is the intrinsic delay) of a selected cell is measured across different MLS and gate voltages (V\textsubscript{WL}), as shown in Fig.~\ref{fig:mls_calib}(b-c), for fast and slow dies, respectively. In these measurements, V\textsubscript{WL} ranging from -0.2~V to 1.8~V in 50~mV steps are applied to evaluate the combined impact of gate bias and MLS on $\Delta t_d$. A notable resolution $\Delta t$ of 100~ps is achieved at 0.65~V. As expected, $\Delta t_d$ resembles the characteristic hysteresis behavior of the FeFET, validating its suitability for delay-based state modulation and calibration.

The impact of MLS calibration on computation correctness is reflected in the timing margin between adjacent time-domain outcomes. Reliable digitization requires that the accumulated delay uncertainty remains well below the measured separation $\Delta s$ between neighboring levels. The total timing uncertainty can be written as
\begin{equation}
\sigma_T^2 = k\,\sigma_{dL}^2 + (N-k)\,\sigma_{dH}^2 + \sigma_{\mathrm{jit}}^2 + \sigma_{\mathrm{TDC}}^2 ,
\label{eq:sigmaT}
\end{equation}
where $N$ is the number of DEs in the chain and $k$ the number of DEs contributing the short-delay state. The terms $\sigma_{dL}$ and $\sigma_{dH}$ denote the stage-to-stage dispersion of the short- and long-delay states, respectively, while $\sigma_{\mathrm{jit}}$ and $\sigma_{\mathrm{TDC}}$ represent other jitter sources (\textit{e.g.}, supply noise, thermal noise) and TDC uncertainty. Without calibration, $\sigma_{dL}$ is dominated by intrinsic D2D mismatch, which accumulates with $k$ and reduces the margin between adjacent delay levels. With MLS calibration, the residual D2D dispersion of $t_{dL}$ is bounded by the tuning resolution,
\begin{equation}
\sigma_{dL,\mathrm{D2D}}^{\mathrm{calibrated}} \le \frac{\Delta {t}}{\sqrt{12}},
\qquad \Delta {t}\le 100~\mathrm{ps},
\label{eq:sigma_dL_cal}
\end{equation}
thereby replacing the process-limited D2D spread with a resolution-limited bound. In contrast, $t_{dH}$ is stabilized by the leaker path, which clamps its delay and limits sensitivity to FeFET off-state variability.

Assuming midpoint decision thresholds and Gaussian timing uncertainty, the probability of misclassification between adjacent outcomes is
\begin{equation}
P_{\mathrm{err}} = Q\!\left(\frac{\Delta s}{2\sigma_T}\right),
\label{eq:Perr}
\end{equation}
where $Q(\cdot)$ is the complementary Gaussian cumulative distribution function. For a fixed $\Delta s$, the impact of calibration on computation correctness is captured by the ratio of timing uncertainties,
\begin{equation}
\frac{\sigma_T^{\mathrm{uncalibrated}}}{\sigma_T^{\mathrm{calibrated}}}
\approx
\sqrt{
\frac{
k\,(\sigma_{dL,\mathrm{D2D}}^{\mathrm{uncalibrated}})^2 + \Phi
}{
k\,(\Delta {t}/\sqrt{12})^2 + \Phi
}},
\label{eq:margin_gain}
\end{equation}
where $\Phi=(N-k)\sigma_{dH}^2+\sigma_{\mathrm{jit}}^2+\sigma_{\mathrm{TDC}}^2$ groups all non-calibrated contributions. Since $\sigma_{dL,\mathrm{D2D}}^{\mathrm{uncalibrated}}$ dominates prior to calibration, MLS tuning increases the timing margin by directly suppressing the leading D2D term. This margin restoration exponentially reduces $P_{\mathrm{err}}$ via \eqref{eq:Perr}, explaining the experimentally observed separation of MAC and logic delay levels and establishing the accuracy gain required for reliable operation under variability when scaling to larger arrays.

\section{\rev{Experimental Results}}
\label{sec:results}
\subsection{\rev{Experimental Results of MAC and Logic Operations}}
\label{sec:mac_logic_results}
\begin{figure}[!t]
    \centering
    \includegraphics[width=0.7\columnwidth,angle=-90,trim={2.4cm 0.6cm 0.6cm 0.6cm},clip]{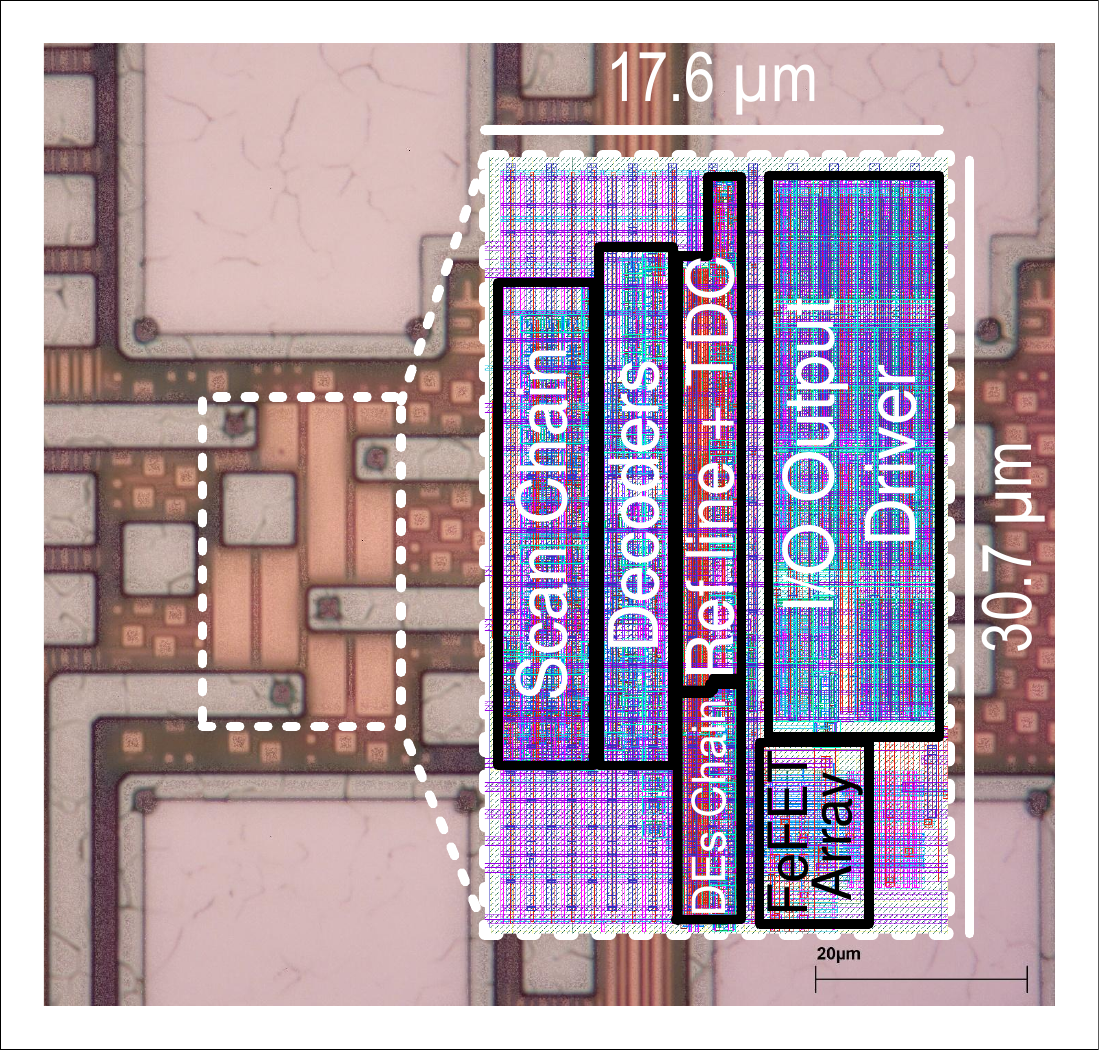}
    \caption{Optical micrograph and layout of the proposed TD-nvIMC macro. The area of the PoC including the DE chain, TDC and reference line, FeFET CAM array, scan chain, and I/O is $17.6\times 30.7~\mu m^2$.}
    \label{fig:chip}
\end{figure}

\begin{figure}[!t]
    \centering
    \includegraphics[width=\columnwidth,trim={0.6cm 0.6cm 0.6cm 0.6cm},clip]{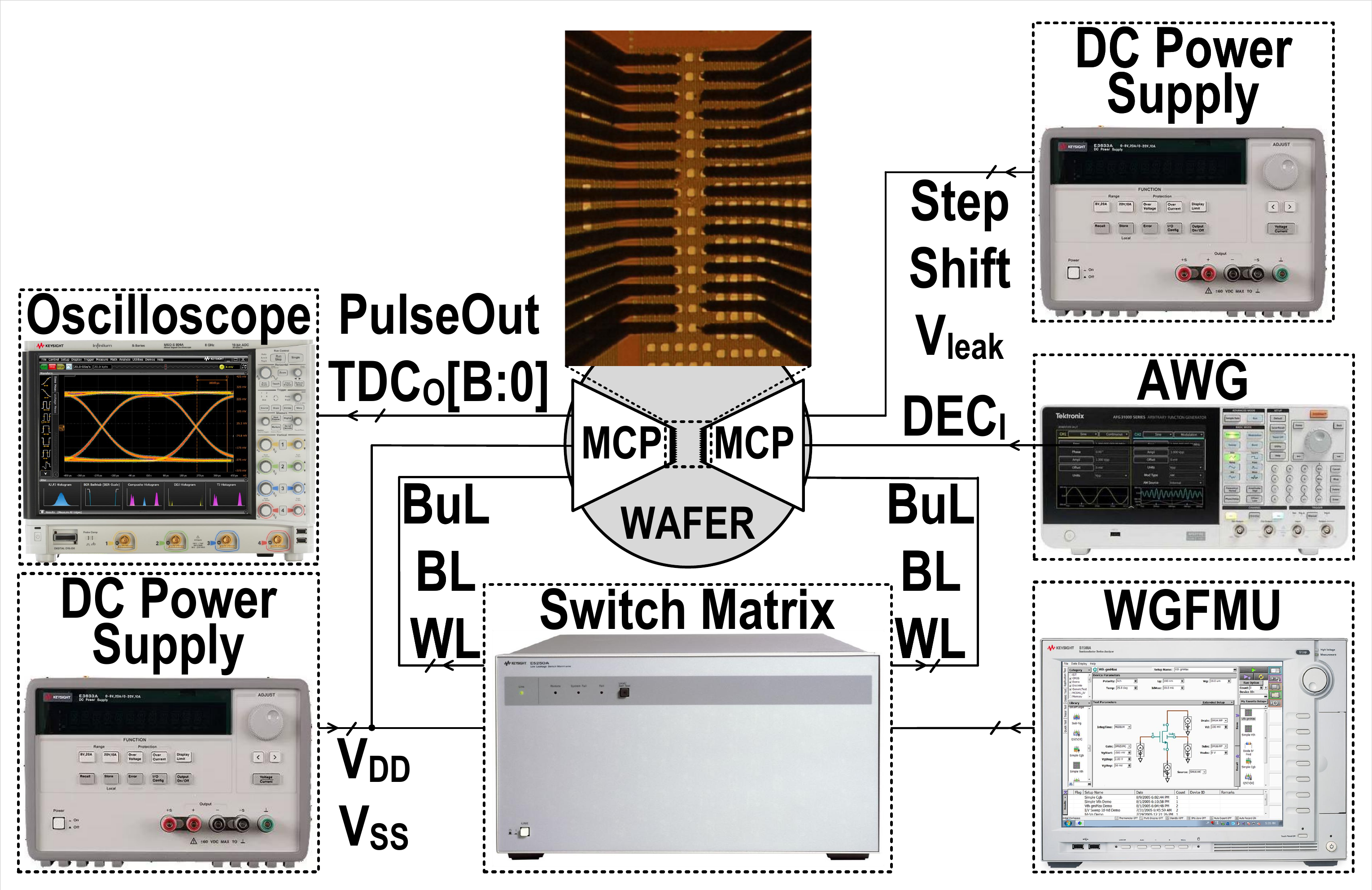}
    \caption{Experimental setup. Two MCPs with 15 probes each are used for on-die probing. The IO driver output is sampled using a high-speed oscilloscope with $50~\Omega$ input. The control signals and input pulse are generated by AWGs\rev{, write pulses by the WGFMU,} and power supplies are used for $V_{DD}$ and bias signals such as $V_{leak}$, \textit{step}, and \textit{shift}.}
    \label{fig:setup}
\end{figure}

The fabricated proof-of-concept (PoC) includes a 3$\times$3 CAM array, a 3-stage DE chain, a 2-bit TDC, a reference delay line, I/O driver, decoder, and testing circuits, and is powered by a 0.85-V supply ($V_{DD}$). An optical micrograph of the fabricated PoC is shown in Fig.~\ref{fig:chip}. The PoC macro area is $17.6 \times 30.7~\mu m^2$. Note that the I/O occupies a significant area while the CAM array, DE chain, and TDC occupy a relatively compact footprint. 
The experimental setup is illustrated in Fig.~\ref{fig:setup}. On-die measurements are performed using two multicontact probes (MCPs) with 15 probes each. The I/O driver output is sampled using a high-speed oscilloscope with $50~\Omega$ input. \rev{DC supplies provide the supply voltage ($V_{DD}$) and bias signals, while the control signals and input pulse are generated by an arbitrary waveform generator (AWG). Furthermore, a waveform generator/fast measurement unit (WGFMU) is used to generate the program and erase pulses for the FeFETs and to perform current measurements for device characterization.}

\begin{figure}[!t]
    \centering
    \includegraphics[width=\columnwidth,trim={0.6cm 0.6cm 0.6cm 0.6cm},clip]{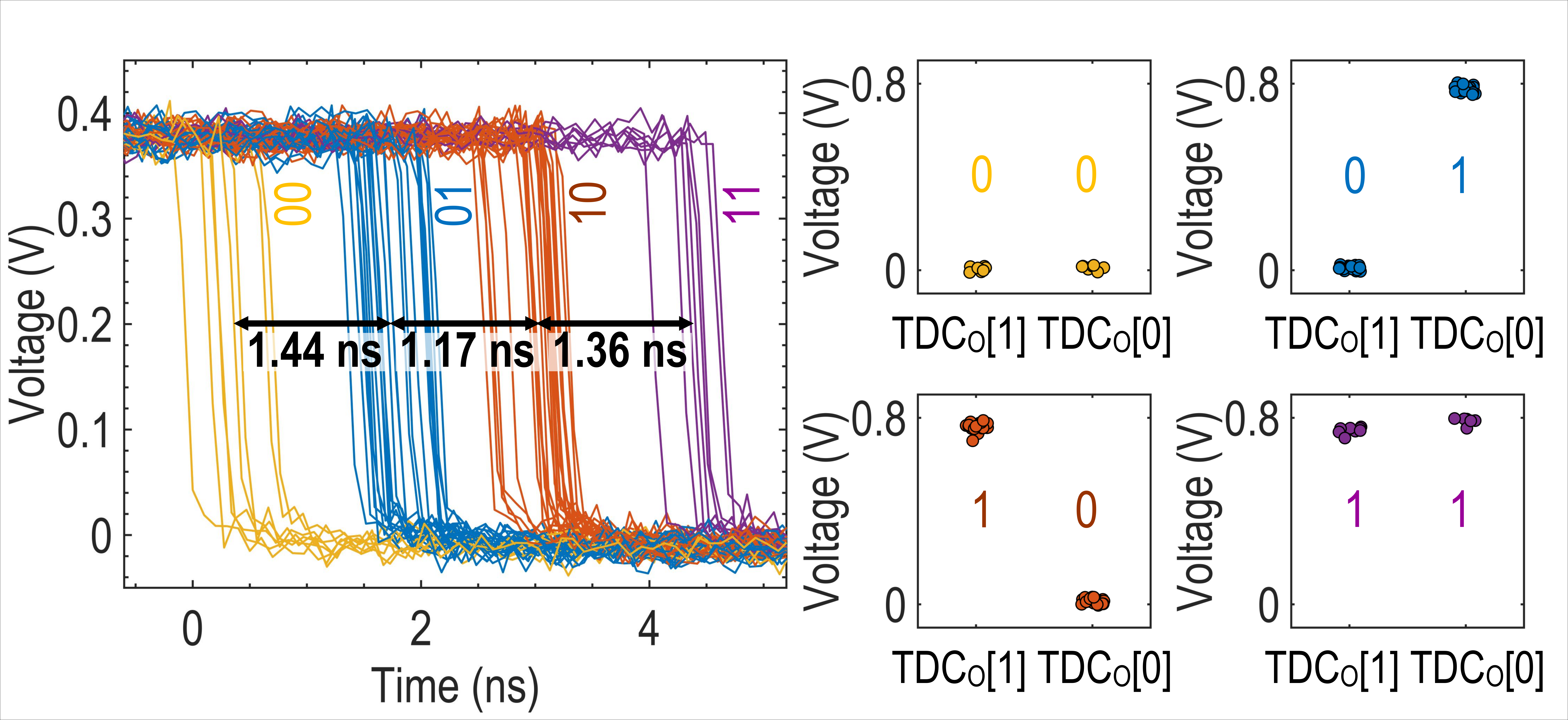}
    \caption{XOR-based MAC experimental results. TDC\textsubscript{O}: 00 (+3), 01 (+1), 10 (-1), 11 (-3). }
    \label{fig:mac_xor_meas}
\end{figure}
XOR-MAC is experimentally validated, covering all $2^6=64$ input combinations of three binary activations-weights pair. Each \rev{configuration} is loaded into the CAM array and triggered through the DE chain. The resulting output pulse (delay) is sampled at $DEC_O$ and digitized using the 2-bit TDC, which maps the delay to a corresponding $TDC_O$ code. The observed step size between successive MAC results is 1.3~ns, with four distinct levels clearly distinguishable as shown in Fig.~\ref{fig:mac_xor_meas}. This confirms the ability of the macro to resolve the full dynamic range of XOR-based MAC operations. 

\begin{figure}[!t]
    \centering
    \includegraphics[width=\columnwidth,trim={0.6cm 0.6cm 0.6cm 0.6cm},clip]{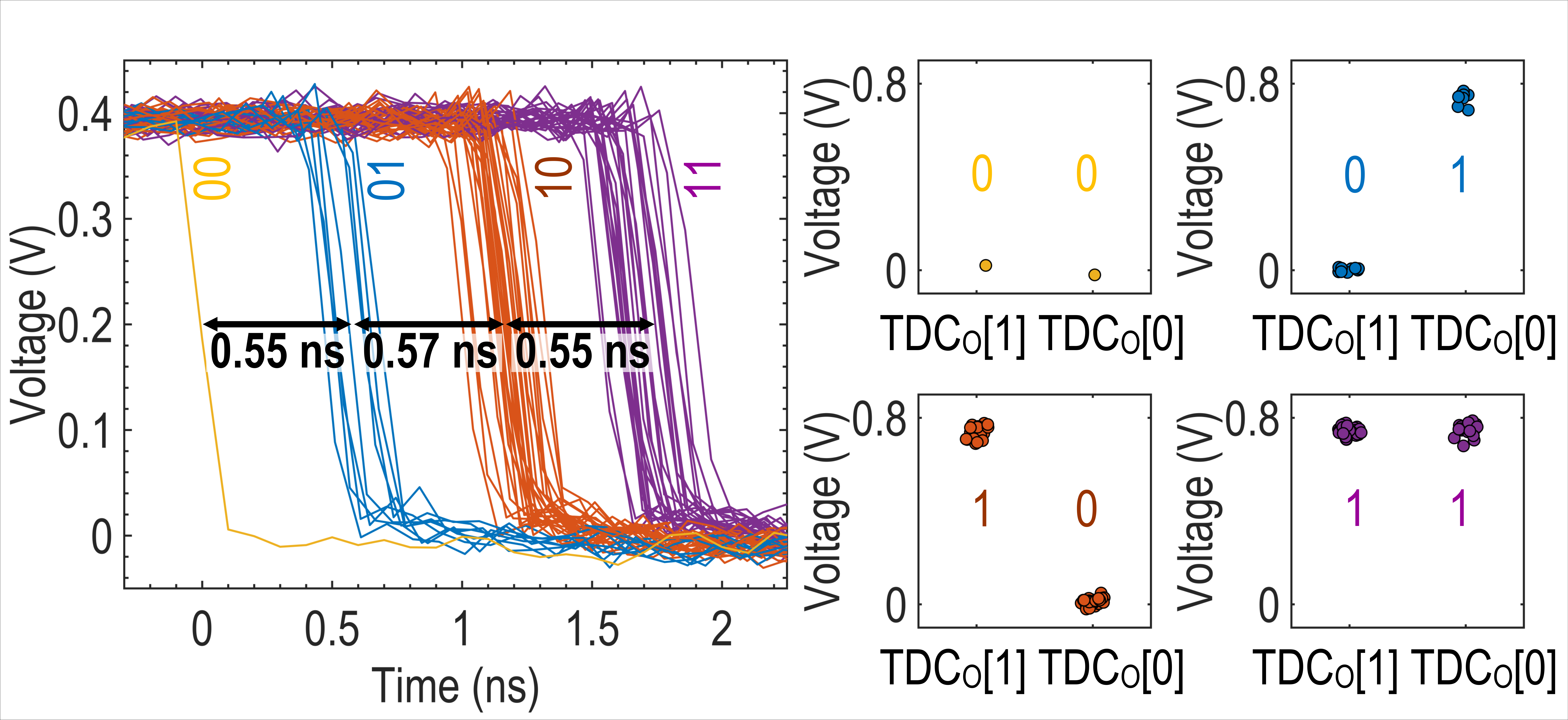}
    \caption{AND-based MAC experimental results.  TDC\textsubscript{O}: 00 (+3), 01 (+2), 10 (+1), 11 (0). }
    \label{fig:mac_and_meas}
\end{figure}

The experimental results also validate AND-based MAC operation under the same test setup. In this mode, the step size is reduced to $\Delta s=550~ps$, as shown in Fig.~\ref{fig:mac_and_meas}. This finer resolution is a direct consequence of the reduced number of FeFETs involved in the operation--only one per stage in AND-MAC versus two in XOR-MAC. Compared to previous works ~\cite{Yin2024,Luo2021}, this $\Delta s$ represents an improvement of over $2000\times$, confirming the advantage of the proposed architecture in both resolution and delay granularity. It can be noted that the measured spread in XOR-MAC delays is noticeably wider, which can be attributed to the involvement of two FeFET devices per CAM cell. This doubles the number of active resistive paths and increases the sensitivity to D2D variations.

\begin{figure}[!t]
    \centering
    \includegraphics[width=\columnwidth,trim={0.6cm 0.6cm 0.6cm 0.6cm},clip]{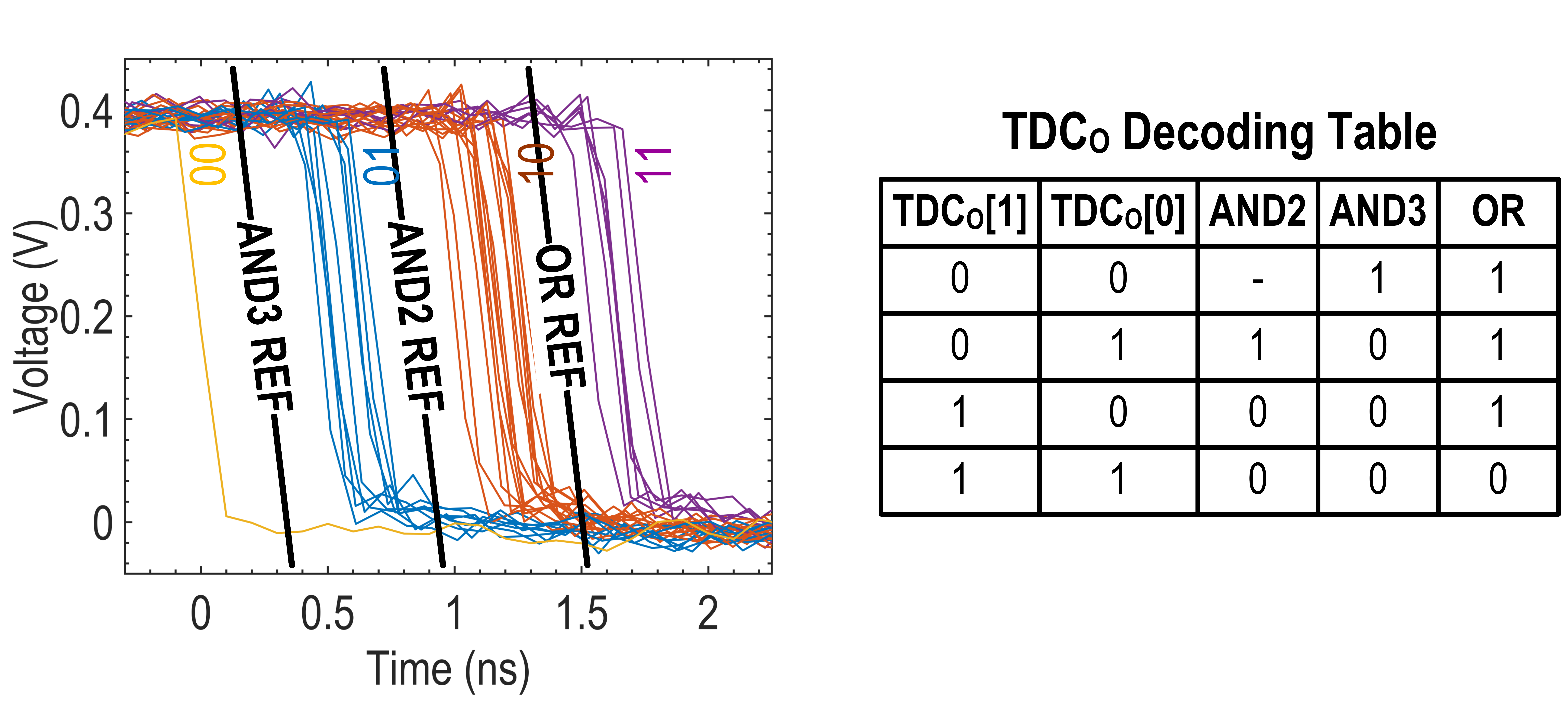}
    \caption{IMC Boolean logic operations for two and three inputs. Truth table and experimental results. }
    \label{fig:logic_meas}
\end{figure}

In-memory logic operations are experimentally validated by performing AND and OR functions directly on the stored states of the FeFET cells. As shown in Fig.~\ref{fig:logic_meas}, the measured $DEC_O$ outputs validate correct logic behavior across various input combinations. For 3-bit inputs, eight possible combinations are tested. For 2-bit cases, all four input combinations are evaluated across three column pairs (\textit{i.e.,} first–second, first–third, and second–third). Each combination is tested twice, with the unselected cell programmed to ‘0’ and ‘1’, resulting in 24 test cases per logic function. The results confirm that $T_D$ encodes the correct logic outcome based on the combination of stored values. For AND, only the all-ones input yields $t_{dL}$ (logic ‘1’), whereas for OR operation, only the all-zeros case results in the longest delay (logic ‘0’).

\subsection{\rev{FeFET Retention and Endurance Characterization}}
\label{sec:reliability}
%\rev{The proposed TD-nvIMC macro relies on stable FeFET $V_T$ states to support the MLS-based calibration in Section~\ref{sec:MLS}. Prior work has reported retention and endurance characteristics of HfO$_2$-based FeFETs in advanced CMOS platforms, including 28-nm technologies~\cite{Raffel2023}. Here, retention was evaluated by tuning devices to multiple MLS and monitoring $I_{BL}$ over $10^5$~s ($\approx 28$~h) at room temperature using read pulses with $V_{WL}=0.6$~V and $V_{BL}=0.8$~V. As shown in Fig.~\ref{fig:reliability}(a), the measured currents exhibit negligible drift across the tested states. This stability is most critical for the LVT-programmed levels that directly determine $t_{dL}$, while potential drift in the HVT state is further desensitized by the parallel leaker path (Fig.~\ref{fig:arch}), which clamps $t_{dH}$ against minor off-state variations.}

\begin{figure}[!t]
    \centering
    % Subfigure A
    \begin{subfigure}[b]{0.49\columnwidth}
        \centering
        % TRIM ADJUSTMENT: Change {left bottom right top} values to crop whitespace
        \includegraphics[width=1.1\linewidth, trim={0.6cm 0.6cm 0.5cm 0.6cm}, clip]{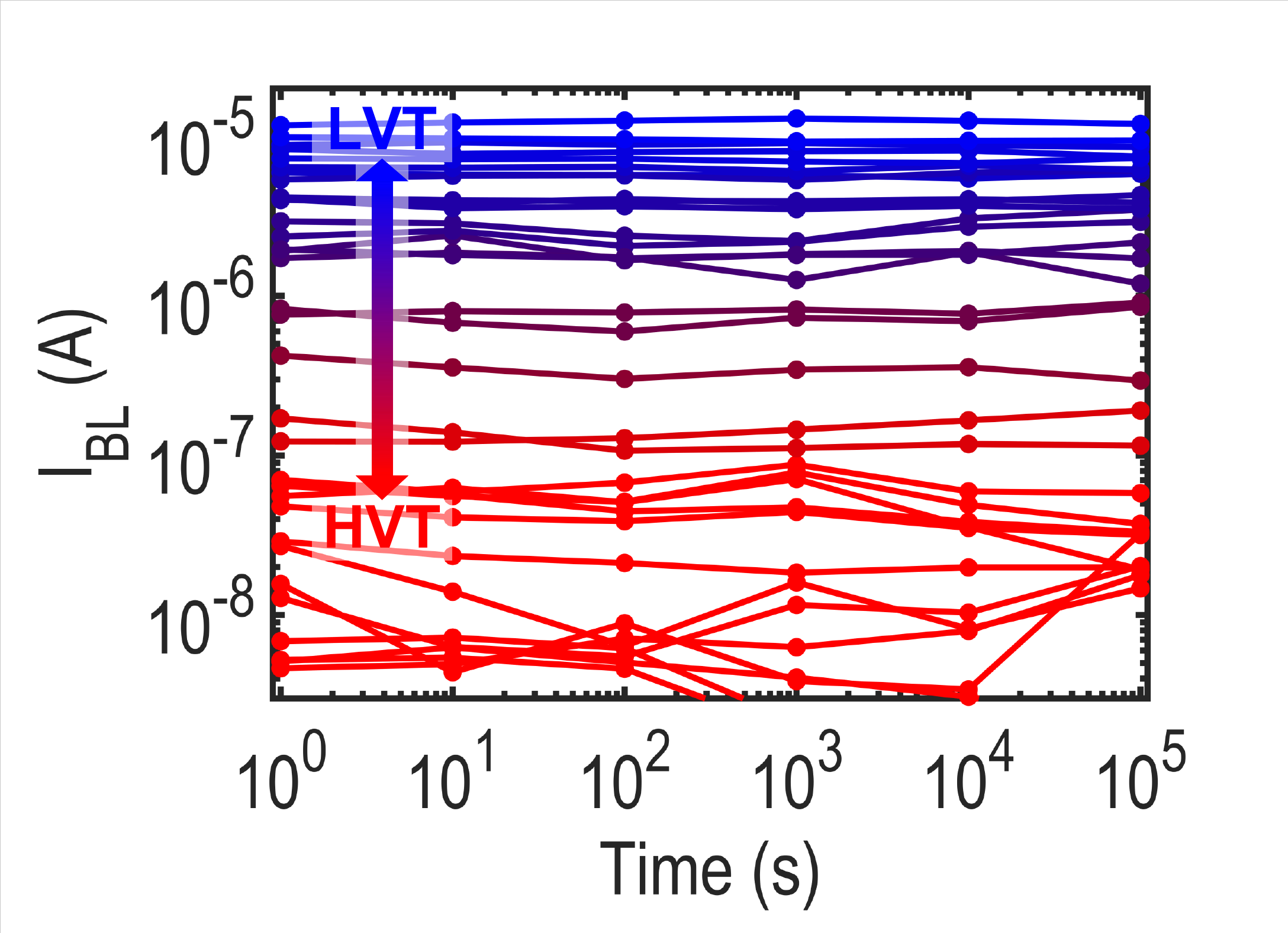}
        \caption{}
    \end{subfigure}%
    \hfill%
    % Subfigure B
    \begin{subfigure}[b]{0.49\columnwidth}
        \centering
        \includegraphics[width=1.1\linewidth, trim={0.6cm 0.6cm 0.5cm 0.6cm}, clip]{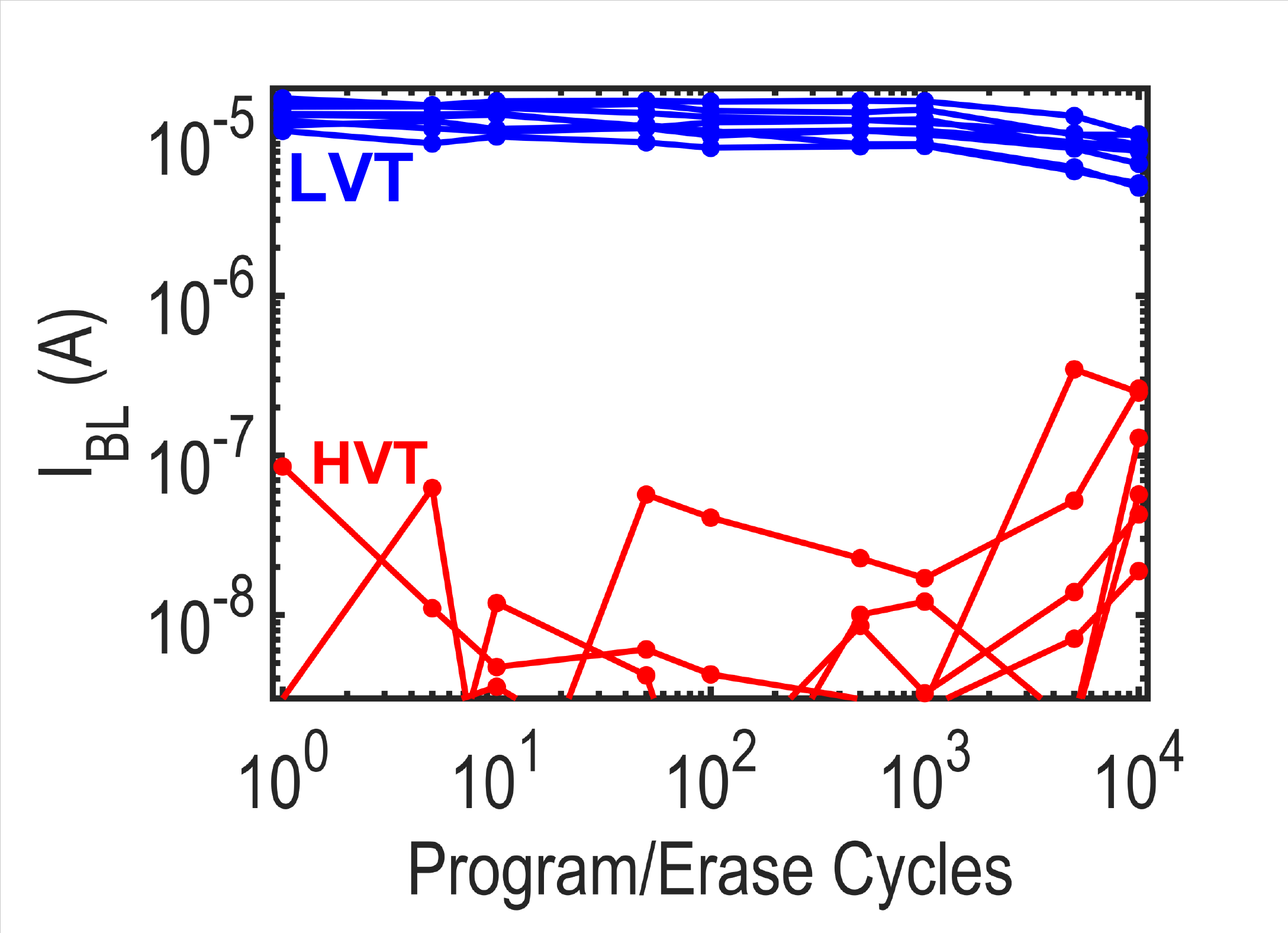}
        \caption{}
    \end{subfigure}
    \caption{\rev{FeFET reliability characterization. (a) Retention: measured $I_{BL}$ over time for multiple MLS, showing stable readout up to $10^5$~s. (b) Endurance: measured $I_{BL}$ versus full-swing program/erase cycles.}}
    \label{fig:reliability}
\end{figure}

\rev{The proposed TD-nvIMC macro relies on stable FeFET $V_T$ states to support the MLS-based calibration in Section~\ref{sec:MLS}. Prior work on HfO$_2$-based FeFET arrays in 28-nm CMOS reports stable retention up to $10^4$~s at 85$^\circ$C and endurance on the order of $10^3$--$10^4$ full-swing cycling before stronger degradation at higher stress~\cite{Raffel2023}. In Addition, multilevel FeFET studies report that intermediate states can be as stable as the saturated states~\cite{Halid2023,Liu2022}. Here, retention was evaluated by tuning devices to multiple MLS and monitoring $I_{BL}$ over $10^5$~s ($\approx 28$~h) at room temperature using read pulses with $V_{WL}=0.6$~V and $V_{BL}=0.8$~V. As shown in Fig.~\ref{fig:reliability}(a), the measured currents exhibit negligible drift across the tested states. This stability is most critical for the LVT-programmed levels that directly determine $t_{dL}$, while potential drift in the HVT state is further desensitized by the parallel leaker path (Fig.~\ref{fig:arch}), which clamps $t_{dH}$ against minor off-state variations.}

\rev{Endurance is characterized under repeated full program/erase (P/E) cycling to provide a conservative reliability bound under worst-case switching stress. The measured $I_{BL}$ as a function of P/E cycles using full-swing pulses is shown in Fig.~\ref{fig:reliability}(b). The read current remains consistent over the initial cycling regime (up to $\sim 10^3$ cycles),  while a gradual shift is observed as cycling progresses into the several-thousand-cycle range, which is expected under aggressive full-swing P/E stress. Importantly, the proposed calibration flow is less stressful than this measurement: it starts from a single program operation and then applies controlled bulk-assisted partial erase adjustments. Prior work has reported that partial switching can improve endurance compared to repeated full-swing cycling~\cite{Duan2022}, consistent with reports that bipolar switching accelerates degradation through charge trapping and interface degradation under repeated program/erase switching stress~\cite{Dominik2024}. Therefore, the effective endurance margin in normal calibrated operation is expected to be higher than what is implied by worst-case full-swing cycling.}

\section{Comparison}
\label{sec:comparison}
A comparison between the proposed macro and previous TD-nvIMC implementations is listed in Table~\ref{tab:comparison_nvIMC}, highlighting improvements in integration level, $\Delta s$, calibration resolution, per-cell throughput, and area and energy efficiency. The proposed macro, fabricated in a 28~nm CMOS process, demonstrates for the first time a fully integrated FeFET-based TD-nvIMC system that includes the memory array, DE chain, and TDC. Previous works, such as~\cite{Luo2021,Yin2024}, report only small-scale demonstrations using discrete FeFET devices for delay modulation, without full macro-level integration. These implementations operate on stand-alone components rather than structured arrays, avoiding challenges such as write disturb, interconnect parasitics, and coupling effects that arise in array-level operation and impact delay uniformity and robustness.

A notable distinction lies in the temporal resolution achieved by this work. A $\Delta_s$ of 550~ps is demonstrated, representing a sub-ns resolution that is nearly $2000\times$ finer than the coarse delay steps exceeding 10~$\mu$s and 1~$\mu$s experimentally reported in~\cite{Luo2021,Yin2024}, respectively. In these studies, large delays are used to experimentally confirm the functional tunability of FeFET-programmed delay elements. However, energy efficiency and throughput metrics are reported based on circuit-level simulations that assume significantly smaller delay steps, often more than three orders of magnitude below the experimentally verified values, leading to extrapolated figures like 51,318~TOPS/W and 8,563~TOPS/W. These simulation-based estimates do not account for key non-idealities, such as variability and parasitic effects, which are increasingly critical at sub-ns scales. 
%In contrast, the present work experimentally demonstrates calibrated delay steps in the sub-ns regime directly on silicon, validating the operation required to support the reported throughput and energy efficiency metrics.
\rev{While circuit simulations of our macro indicate feasible operation at $\Delta s \approx 100$~ps (yielding $>5\times$ higher throughput), experimental measurements show that these non-idealities limit the robust operating point to $\Delta s = 550$~ps; accordingly, the reported performance metrics are anchored to this experimentally validated silicon regime rather than optimistic extrapolation.}
%\rev{Indeed, while circuit simulations of our macro indicate feasible operation at $\Delta s \approx 100$~ps (yielding $>5\times$ higher efficiency), measurements reveal that real-world constraints limit the robust resolution to 550~ps. Consequently, the reported metrics are anchored to this experimentally verified operating point, ensuring that throughput and energy-efficiency figures reflect a rigorously validated hardware regime rather than optimistic extrapolation.}
%\rev{While circuit simulations of the proposed macro suggest feasibility at finer delay steps (e.g., $\Delta s \approx 100$~ps), which would imply $>5\times$ higher efficiency, the reported performance metrics are based strictly on the experimentally validated silicon operating point of $\Delta s = 550$~ps, ensuring that throughput and energy-efficiency figures reflect a rigorously validated hardware regime rather than optimistic extrapolation.}
%\rev{While circuit simulations of our macro indicate feasible operation at $\Delta s \approx 100$~ps (yielding $>5\times$ higher efficiency), experimental measurements show that real-world effects, including D2D variability, timing jitter, and measurement noise, limit the robust operating point to $\Delta s = 550$~ps; accordingly, the reported performance metrics are based strictly on this experimentally validated silicon regime rather than optimistic extrapolation.}

To quantify computational density, this work reports area efficiency (in TOPS/mm$^2$), defined as total throughput normalized by the occupied area. The proposed design achieves 3.7 TOPS/mm$^2$, nearly an order of magnitude higher than the ReRAM-based design in~\cite{Hung2023}, which reports 0.387 TOPS/mm$^2$. Compared to the STT-MRAM design in~\cite{Seungchul2022}, which reaches 4.43 TOPS/mm$^2$, our area efficiency is lower; however, this difference arises primarily from design scale and layout considerations. Our implementation is a PoC that integrates a 3×3 FeFET array alongside debug and DFT circuits, which occupy 67\% of the total chip area. As the array scales, the proportion of area devoted to computation will increase substantially, and the overall area efficiency is expected to improve accordingly.

Since some works do not report area due to being simulation-only or implementing partial hardware (\textit{e.g.}, memory-only in~\cite{Saito2021}), we also report throughput normalized by the number of active memory cells, resulting in a measured 222.2 MOPS/cell. This metric enables fair comparison across architectures of varying scales and directly reflects the experimentally validated sub-ns delay steps that support the reported energy efficiency of 1887 TOPS/W. These performance figures are based on measured silicon results and are specifically attributed to the active compute path, including the DE chain, TDC, and RDL, while excluding DFT and peripheral logic. In contrast, previous FeFET-based time-domain studies report throughput and energy efficiency values derived solely from simulations that assume fine delay steps, often several orders of magnitude smaller than experimentally demonstrated, without providing calibrated or validated measurements on chip.

For instance, \cite{Saito2021} demonstrates only the fabrication and silicon validation of a FeFET-based 1T1R memory array, where analog vector-matrix multiplications are performed via current summation along BLs. However, the reported energy efficiency and inference performance rely entirely on simulations incorporating capacitor-based ADCs and digital accumulation stages, without hardware validation of these critical components. In contrast, the ReRAM-based macro presented in~\cite{Hung2023} implements a complete 8~Mb array with time-domain ADC-based readout, providing a fully integrated and silicon-verified compute pipeline. Nevertheless, its core computation depends on analog current summation, and thus remains susceptible to small-signal margins and parasitic effects inherent in current-mode crossbar architectures. The approach presented in this work eliminates these analog accumulation challenges by directly quantizing computation results in the time domain. Sub-ns delay steps, calibrated and verified on silicon, are leveraged to perform in-memory MAC operations with high resolution and robustness. This enables a fully digital-compatible pathway for integrating energy-efficient nvIMC into standard design flows.

\begin{table*}[t]
    \centering
    \renewcommand{\arraystretch}{1.2}
    \setlength{\tabcolsep}{4pt} 
    \caption{\textsc{{Comparison with State-of-the-Art nvIMC Architectures}}}
    \label{tab:comparison_nvIMC}
    \begin{tabular}{lC{1.5cm}C{1.8cm}C{1.8cm}C{1.8cm}C{1.8cm}C{1.8cm}}
        % \hline
        \textbf{Reference} & \textbf{This Work}& IEDM’21 \cite{Luo2021} & TCAD’24 \cite{Yin2024} & VLSI’21 \cite{Saito2021} & Nat.’22 \cite{Seungchul2022} & JSSC’23 \cite{Hung2023} \\ \hline
        \hline 
       \textbf{Technology (nm)} & 28 & 14 & 40 & 22 & \rev{28} & \rev{22} \\ \hline
       \textbf{NVM Type} & FeFET & FeFET & FeFET & FeFET & STT-MRAM & ReRAM \\ \hline
        \textbf{Integration Level} & Fully integrated PoC & DEs only & Discrete elements DEs only & Memory array only & Fully integrated & Fully integrated \\ \hline
        \textbf{Signal Domain} & Time & Time & Time & Voltage & Voltage (TD readout) & Voltage (TD readout) \\ \hline
       \textbf{Measured} $\Delta s$ & 550 ps & $>$10 $\mu$s & $>$1 $\mu$s & - & - & - \\ \hline
        \textbf{Calibration Resolution} & 100 ps & No calibration & No calibration & - & - & - \\ \hline
       
        \textbf{Area efficiency (TOPS/mm$^2$)} & 3.70 & - & - & - & 4.43 & 0.387 \\ \hline
        \textbf{Throughput per Cell (MOPS/cell)} & 222.22 & - & 29.3\textsuperscript{(a)} & 56.96\textsuperscript{(a)} & - & 0.64 \\ \hline
        \textbf{Energy Efficiency (TOPS/W)} & 1887\textsuperscript{(b)} & 51318\textsuperscript{(a)} & 8563\textsuperscript{(a)} & 2200\textsuperscript{(a)} & 405 & 416.5 \\ \hline
        \hline
        \multicolumn{7}{l}{\textsuperscript{(a)} From simulations.}\\
        \multicolumn{7}{l}{\textsuperscript{(b)} Includes power of TDC, reference line, and decoders.} \\
    \end{tabular}

\end{table*}

% \vspace{-0.5em}
\section{Conclusion}
\label{sec:conclusion}
This work demonstrates a reconfigurable FeFET-based TD-nvIMC macro  that integrates MLS calibration to enhance delay precision and robustness. The proposed architecture supports both binary MAC operations and in-memory logic functions within the same memory structure, showcasing the versatility of the design. Experimental results highlight a fine delay resolution of $\Delta t \leq 100~\text{ps}$ and a delay step size of $\Delta s \simeq 550~\text{ps}$, representing more than a 2000$\times$ improvement over prior work. These results demonstrate the effectiveness of the calibration scheme in mitigating delay mismatches and device variations. Overall, the combination of reconfigurability, high throughput, and precise time-domain control makes this approach highly suitable for next-generation IMC platforms, particularly in applications demanding energy-efficient and parallel processing such as edge AI, neural network acceleration, and low-power embedded systems.

\section*{Acknowledgment}

The authors thank Yair Keller and Ilan Lipschutz for their support during the experimental measurements and Prof. Ariel Cohen for the useful discussions.

\ifCLASSOPTIONcaptionsoff
  \newpage
\fi

\bibliographystyle{ieeetr}
\bibliography{references}

\end{document}